\documentclass[prd,twocolumn,nofootinbib,nobibnotes,twoside,]{revtex4-1}
\usepackage{graphicx}
\usepackage{amssymb}
\usepackage{amsmath}
\usepackage{color}
\usepackage{float}
\usepackage{ulem}
\usepackage{accents}
\usepackage{graphicx}
\usepackage{graphicx}
\usepackage{amsfonts}
\usepackage[colorlinks=true,
pdfstartview=FitV,linkcolor=blue,
citecolor=blue,urlcolor=blue,breaklinks=true]
{hyperref}
\usepackage{array}
\usepackage{float}
\usepackage{placeins}
\usepackage[dvipsnames]{xcolor}
\usepackage{csquotes}
\usepackage{bbold}
\usepackage{units}
\newcommand{\orcid}[1]{\href{https://orcid.org/#1}{\includegraphics[width=10pt]{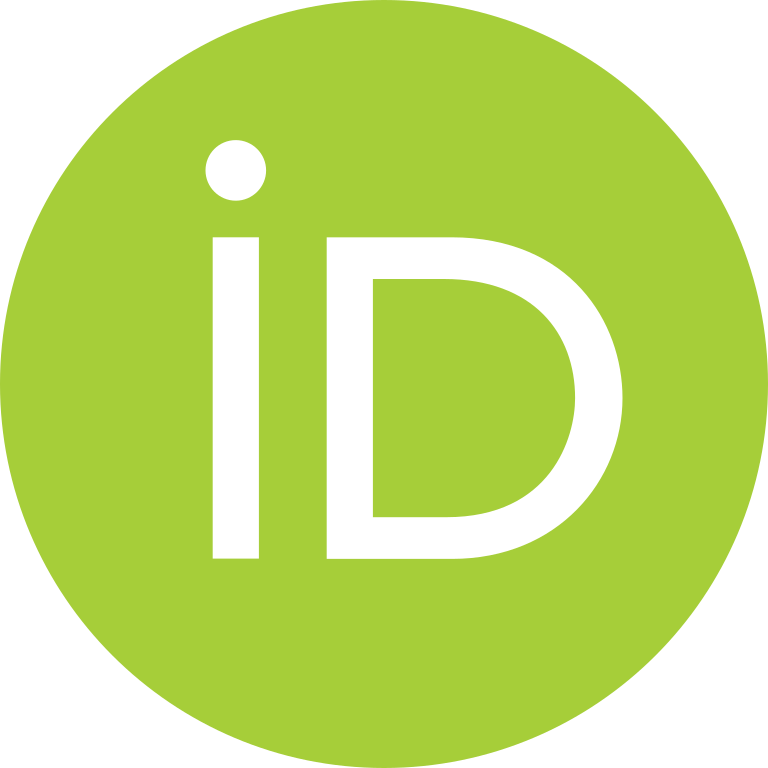}}}

\usepackage{enumitem}

\usepackage{pbox}

\usepackage{fancyhdr}

\fancypagestyle{alim}{\fancyhf{}\fancyhead[C]{{\large PHYSICAL REVIEW D {\bf{102}}, 076001 (2020)}}\fancyfoot[C]{\thepage}\fancyfoot[R]{\footnotesize{Published by the American Physical Society}}}

\newcommand\blfootnote[1]{%
  \begingroup
  \renewcommand\thefootnote{}\footnote{#1}%
  \addtocounter{footnote}{-1}%
  \endgroup
}

\newcolumntype{C}[1]{>{\centering\arraybackslash}m{#1}}

\renewcommand{\eqref}[1]{\mbox{Eq.~(\ref{#1})}}

\definecolor{ForestGreen}{rgb}{0.13,0.55,0.13}

\usepackage{setspace}

\begin{document}
\setstretch{1.05}

\title{Magnetic-conductivity effects on electromagnetic propagation \\ in dispersive matter}

\author{Pedro D.S. Silva\orcid{0000-0001-6215-8186}$^a$}
\email{pedro.dss@discente.ufma.br}
\affiliation{$^a$Departamento de F\'{i}sica, Universidade Federal do Maranh\~{a}o, Campus
Universit\'{a}rio do Bacanga, S\~{a}o Lu\'is (MA), 65080-805, Brazil}
\author{Manoel M. Ferreira Jr.\orcid{0000-0002-4691-8090}$^a$}
\email{manojr.ufma@gmail.com}
\affiliation{$^a$Departamento de F\'{i}sica, Universidade Federal do Maranh\~{a}o, Campus
Universit\'{a}rio do Bacanga, S\~{a}o Lu\'is (MA), 65080-805, Brazil}
\author{Marco Schreck\orcid{0000-0001-6585-4144}$^a$}
\email{marco.schreck@ufma.br}
\affiliation{$^a$Departamento de F\'{i}sica, Universidade Federal do Maranh\~{a}o, Campus
Universit\'{a}rio do Bacanga, S\~{a}o Lu\'is (MA), 65080-805, Brazil}
\author{Luis F. Urrutia\orcid{0000-0001-9792-7344}$^b$}
\email{urrutia@nucleares.unam.mx}
\affiliation{$^b$Instituto de Ciencias Nucleares, Universidad Nacional Aut\'onoma de M\'exico, \\
04510 M\'exico, Distrito Federal, Mexico\looseness=-1}

\author{(Received 8 June 2020; accepted 31 August 2020; published 2 October 2020)}

\begin{abstract}
The chiral magnetic effect (CME) has been investigated as a new transport phenomenon in condensed
matter. Such an effect appears in systems with chiral fermions and involves an electric current
generated by a magnetic field by means of an ``exotic'' magnetic conductivity. This effect can also
be connected with extensions of the usual Ohm's law either in  magnetohydrodynamics or in
Lorentz-violating scenarios. In this work, we study the classical propagation of electromagnetic
waves in isotropic dispersive matter subject to a generalized Ohm's law. The latter involves currents
linear in the magnetic field and implies scenarios inducing parity violation. We pay special attention
to the case of a vanishing electric conductivity. For a diagonal magnetic conductivity, which includes
the CME, the refractive index is modified such that it implies birefringence. For a nondiagonal magnetic
conductivity, modified refractive indices exhibiting imaginary parts occur ascribing a conducting
behavior to a usual dielectric medium. Our findings provide new insights into typical material properties
associated with a magnetic conductivity.
\begin{itemize}[font=\normalfont]
\item[] 
 \item[DOI:] \href{https://doi.org/10.1103/PhysRevD.102.076001}{10.1103/PhysRevD.102.076001}
\end{itemize}
\end{abstract}

\maketitle

\thispagestyle{plain}
\thispagestyle{alim}

\section{\label{section1}INTRODUCTION}\blfootnote{\vspace{0.2cm}} \blfootnote{\pbox[h]{.465\textwidth}{{{\textit{{\small {Published by the American Physical Society under the terms of the \href{https://creativecommons.org/licenses/by/4.0/}{Creative Commons Attribution 4.0 International} license. Further distribution of this work must maintain attribution to the author(s) and the published article's title, journal citation, and DOI. Funded by SCOAP$^3$.}}}}}}}The chiral magnetic effect (CME) is the macroscopic generation of an electric
current in the presence of a magnetic field as the result of an asymmetry
between the number density of left- and right-handed chiral fermions. It leads to a
current that is linear in the magnetic field~\cite{Kharzeev1}.
This quantum effect has been the subject of extensive research in
particle and field theory as well as nuclear and condensed matter physics. It was
investigated in quark-gluon plasmas with a chiral chemical potential under
the influence of an external magnetic field \cite{Fukushima,Gabriele} and was
also derived in the context of high-energy physics by Vilenkin (in the
1980s) \cite{Schober,Vilenkin} who supposed an imbalance of fermion chirality
in the presence of cosmic magnetic fields in the early Universe. The CME was
studied in cosmology~\cite{Maxim}, as well, where it was applied to explain
the origin of the very high magnetic field strengths (up to $\unit[10^{15}]{G}$)
observed in neutron stars \cite{Leite,Dvornikov}. An interesting question concerns the
possible impact of the external axial-vector field $V_{5}^{\mu}$ on the
magnitude of the anomalous current in the CME~\cite{Bubnov}, which was also
examined for the polarization tensor of a photon in a fermion plasma under the
influence of $V_{5}^{\mu}$ \cite{Akamatsu}. Connections between the CME and matter subject to
the electroweak interaction were established, too \cite{Maxim1, Maxim2}.

In condensed matter systems, the CME plays a most important role.
It appears as a relevant effect in Weyl semimetals, where it is usually
connected to the chiral anomaly associated with Weyl nodal points \cite{Burkov}.
In such materials, massless fermions acquire a drift velocity along
the magnetic field, whose direction is given by their chirality.
Opposite chirality implies opposite velocities, creating a chiral-fermion
imbalance that is proportional to the chiral magnetic current. The first
experimental observation of the CME was reported in 2014 \cite{Li}. Several
investigations have been carried out on the properties of this phenomenon,
including the CME in the absence of Weyl nodes \cite{Chang}, anisotropic
effects stemming from tilted Weyl cones \cite{Wurff}, the CME and anomalous
transport in Weyl semimetals \cite{Landsteiner}, quantum oscillations
arising from the CME \cite{Kaushik}, computation of the electromagnetic
fields produced by an electric charge near a topological Weyl semimetal with
two Weyl nodes \cite{Ruiz}, and chiral superconductivity \cite{Kharzeev2}.

An interesting relation between the CME and Maxwell-Carroll-Field-Jackiw
electrodynamics has been established in the literature \cite{Qiu} by examining
the connection between the CME and Lorentz-violating theories including axion
electrodynamics. The possibility of Lorentz invariance violation was
proposed in the context of physics at the Planck scale such as strings~\cite{Kostelecky:1988zi}.
Presently, the Standard Model extension \cite{Colladay},
where fixed background tensor fields are coupled to the dynamical
fields, is usually employed to parametrize it.
A violation in the photon sector can occur by means of a $CPT$-odd
\cite{CFJ} or a $CPT$-even term \cite{KM}. A Lorentz-violating extension for the
current density that resembles the macroscopic description of the CME can be
found in Ref.~\cite{Bailey}, where some preliminary studies
of aspects of Lorentz-violating electrodynamics in continuous matter were performed.

Additional motivation for getting more insight into conduction currents driven by magnetic fields comes
from magnetohydrodynamics. After making some simplifying assumptions for plasmas of colliding
particles, the resistive Ohm's law for the one-fluid model is written as~\cite{Gurnett}
\begin{equation}
(\mathbf{E}+\mathbf{V} \times \mathbf{B})^i=\eta^{ij} J^j\,.
\label{ROL}
\end{equation}
Here $\mathbf{V}$ is the average velocity of the electrons and ions, while $\mathbf{J}$
is the total current density. The expression $\eta^{ij} J^j $ corresponds to an effective
collision term for electrons and ions and the resistivity $\eta^{ij}$ can, in general, be
a tensor. Even if Eq.~(\ref{ROL}) is not a rigorous model, it is widely used, because it
captures the most important deviations from the ideal magnetohydrodynamic model. Inverting
Eq.~(\ref{ROL}) yields $J^i= \sigma_{ij} E^j+ \sigma^B_{ij} B^j $, with  $\sigma_{ij}$
being the inverse of $\eta^{ij}$ and $\sigma^B_{ij}= \sigma_{ip}V_q\epsilon_{pqj}$
(where $\epsilon_{ijk}$ is the three-dimensional Levi-Civita symbol)
defining a magnetic conductivity. In the following, we will generalize this motivation by
considering the magnetic  conductivity tensor $\sigma^B_{ij}$ to be completely independent
of the electric conductivity tensor $\sigma_{ij}$.

The main purpose of this work is to analyze the possible phenomena of a
magnetic conductivity in a usual dielectric medium. In particular, we are
interested in investigating the effects originating from a magnetic
conductivity on the propagation of electromagnetic waves in a dispersive
dielectric continuous medium characterized by the parameters $\epsilon$
(electric permittivity), $\mu$ (magnetic permeability), and $\sigma$ (Ohmic
conductivity). In this sense, we propose some particular configurations for the magnetic
conductivity to be explored.

We start from the Maxwell equations in a continuous medium,
supplemented by the constitutive relations $\mathbf{D}={\epsilon }\mathbf{E}%
, $ $\mathbf{B}={\mu }\mathbf{H},$ and the magnetic current,
$J_{\mathrm{CME}}^{i}={\sigma}_{ij}^{B}B^{j}\mathbf{,}$ arising from an extension of
Ohm's law. We obtain the corresponding refractive indices and electric fields for the
propagating modes in some scenarios. First, we address an isotropic and an
anisotropic diagonal conductivity tensor, configurations which describe the
chiral magnetic current observed in Weyl semimetals. After doing so, we examine
more exotic configurations of antisymmetric and symmetric nondiagonal conductivity
tensors. In what follows, we will take $\epsilon,\mu,\sigma\in\mathbb{R}$.

On the one hand, the isotropic or anisotropic diagonal magnetic conductivity tensors
generate scenarios of birefringent dielectric crystals, described by two different
refractive indices for each wave vector. On the other hand, the nondiagonal anisotropic
magnetic conductivity (symmetric or antisymmetric) creates the remarkable behavior of a
conducting phase in the dielectric substrate.

This work is outlined as follows. In Sec.~\ref{section2} we briefly
review some basic aspects of electrodynamics in matter, focusing on
the refractive index, constitutive relations, the generalized form of Ohm's law,
and the magnetic current. In Sec.~\ref{section3} we discuss the
general effect of a magnetic conductivity on wave propagation in continuous
dielectric matter. We analyze diagonal conductivity tensors and some special
cases of an exotic conductivity. Section~\ref{section4} is dedicated to
deriving the contributions to the charge and current density that result from
the magnetic conductivity and are needed to guarantee the consistency of
Maxwell's equations. Furthermore, in Sec.~\ref{section5} we study what
impact a magnetic conductivity has on certain phenomena such as the skin
effect and reflection. Finally, we summarize our results in Sec.~\ref{section6}.

\section{\label{section2}Basic aspects of electrodynamics in matter}

The electrodynamic properties of a continuous medium are characterized
by its electric permittivity $\epsilon$, magnetic permeability $\mu$, and
Ohmic conductivity $\sigma$. The static and dynamic behavior is described
by the standard Maxwell equations,
\begin{subequations}
\begin{align}
\nabla \cdot \mathbf{D} &=\rho\,, \quad \nabla \times \mathbf{H-\partial _{%
\mathrm{0}}D=J}\,, \label{MINH} \\[1ex]
\nabla \cdot \mathbf{B} &=0\,, \quad \nabla \times \mathbf{E+\partial _{%
\mathrm{0}}B=0}\,. \label{MHOM}
\end{align}
\end{subequations}
The response of the medium to applied electromagnetic fields is measured in
terms of the polarization vector $\mathbf{P}$ and
magnetization vector $\mathbf{M}$, defined by linear
constitutive relations: $\mathbf{P}={%
\epsilon }_{0}{\chi }_{E}\mathbf{E}$ and $\mathbf{M}={\chi}_{M}\mathbf{H}$,
where ${\chi }_{E}$ and ${\chi }_{M}$ are the electric and magnetic
susceptibility, respectively, of an isotropic ponderable medium.
Such relations allow us to define the electric
displacement field $\mathbf{D}$ and the magnetic flux density $\mathbf{B}$:
\begin{subequations}
\begin{align}
\mathbf{D} &={\epsilon }_{0}\mathbf{E}+\mathbf{P}={\epsilon }_{0}(1+{\chi }%
_{E})\mathbf{E}={\epsilon }\mathbf{E}\,,  \label{eq5B} \\[1ex]
\mathbf{B} &={\mu }_{0}\mathbf{H}+{\mu }_{0}\mathbf{M}={\mu }_{0}(1+{\chi }%
_{M})\mathbf{H}={\mu }\mathbf{H}\,.  \label{eq5C}
\end{align}
\end{subequations}
We then rewrite Eqs.~(\ref{eq5B}) and (\ref{eq5C}) as
\begin{equation}
\begin{pmatrix}
\mathbf{D} \\
\mathbf{H}%
\end{pmatrix}%
=%
\begin{pmatrix}
\mathrm{{\epsilon }\mathbb{1}} & 0 \\
0 & \mathrm{\ {{\mu }^{-1}}\mathbb{1}}%
\end{pmatrix}%
\begin{pmatrix}
\mathbf{E} \\
\mathbf{B}%
\end{pmatrix}\,,
\label{eq8}
\end{equation}
with the identity matrix $\mathbb{1}$ in three spatial dimensions.
By using a plane-wave ansatz for the electromagnetic fields, $\mathbf{E}=\mathbf{E}%
_{0}e^{\mathrm{i}(\mathbf{k}\cdot\mathbf{r}-\omega t)}$ and $\mathbf{B}=\mathbf{B}%
_{0}e^{\mathrm{i}(\mathbf{k}\cdot\mathbf{r}-\omega t)}$, the Maxwell equations and $%
\mathbf{J}=\sigma\mathbf{E}$ yield
\begin{equation}  \label{eq10}
\mathbf{k} \times\mathbf{k} \times\mathbf{E}+{\omega}^{2}\mu \bar{\epsilon}%
(\omega) \mathbf{E}=0\,,
\end{equation}
with $\mathbf{k}^{2}={\omega}^{2}{\mu}\bar{\epsilon}(\omega)$, where
\begin{equation}  \label{eq11}
\bar{\epsilon}(\omega)=\epsilon+\mathrm{i}{\frac{{\sigma}}{{\omega}}}\,,
\end{equation}
is the frequency-dependent electric permittivity of the medium. This complex
permittivity leads to a complex refractive index:\footnote{Note that the
refractive index can be a complex function, in general. Thus, instead of
employing the norm $|\mathbf{k}|$ in the definition of the refractive index,
which is a non-negative, real number, we use $+\sqrt{\mathbf{k}^2}$.
Furthermore, we only consider refractive indices with a non-negative real
part, which is indicated explicitly by the plus sign in front of the square
root.}
\begin{subequations}
\begin{equation}
\label{eq13}
\bar{n}=+\frac{\sqrt{\mathbf{k}^2}}{\omega}=\sqrt{\mu\epsilon +\mathrm{i}{\frac{{\mu\sigma}%
}{{\omega}}}}=n{^{\prime}}+\mathrm{i}n{^{\prime\prime}}\,,
\end{equation}
where
\begin{equation}
\label{eq14}
n^{\prime,\prime\prime}=\sqrt{\sqrt{\Upsilon_0^2+\left(\frac{\mu\sigma}{2\omega}\right)^{2}}\pm \Upsilon_0}\,,\quad \Upsilon_0=\frac{\mu\epsilon}{2}\,.
\end{equation}
\end{subequations}
The imaginary part of the refractive index implies a real exponential
factor, $e^{-\omega n^{\prime\prime}(\hat{\mathbf{k}}\cdot\mathbf{r})}$
with $\hat{\mathbf{k}}\equiv\mathbf{k}/\sqrt{\mathbf{k}^2}$, in the
plane-wave solutions for the electromagnetic fields. This term damps the
amplitude of the wave along its propagation through matter and is related
to the absorption coefficient ${\alpha}=2{\omega}%
n^{\prime\prime}$, whose inverse value determines the penetration depth.
Such a scenario is typical for a conducting medium.

In a linear, continuous, ponderable medium, general constitutive relations can
be envisaged as a theoretical possibility \cite{Palash}, which gained great
attention with the advent of topological insulators \cite%
{Urrutia2,Zangwill}. In systems of this kind, it holds that
\begin{equation}
\begin{pmatrix}
\mathbf{D} \\
\mathbf{H}%
\end{pmatrix}%
=%
\begin{pmatrix}
\epsilon \mathrm{\mathbb{1}} & \alpha \mathrm{\mathbb{1}} \\
\beta \mathrm{\mathbb{1}} & \mathrm{\ {{\mu }^{-1}}\mathbb{1}}%
\end{pmatrix}%
\begin{pmatrix}
\mathbf{E} \\
\mathbf{B}%
\end{pmatrix}%
\,,
\end{equation}%
with additional material parameters $\alpha$, $\beta$ that are not independent and
whose sum provides what is known as the ``activity constant'' of a medium.
Such an extension also occurs in a Lorentz-violating anisotropic
electrodynamics~\cite{Bailey} where
\begin{equation}
\begin{pmatrix}
\mathbf{D} \\
\mathbf{H}%
\end{pmatrix}%
=%
\begin{pmatrix}
\epsilon\mathbb{1}+\kappa_{DE} & \kappa_{DB} \\
\kappa_{HE} & {\mu }^{-1}\mathbb{1}+\kappa_{HB} \\
\end{pmatrix}%
\begin{pmatrix}
\mathbf{E} \\
\mathbf{B}%
\end{pmatrix}%
\,,  \label{eq15}
\end{equation}%
with dimensionless $(3\times 3)$ matrices $\kappa_{DE}$, $\kappa_{DB}$, $\kappa_{HE}$, and $\kappa_{HB}$
that are composed of a vacuum part and a matter part. In component form, the latter relations read
\begin{subequations}
\label{eq17A}
\begin{align}
D^{i}& =\left[ \epsilon {\delta }_{ij}+({\kappa}_{DE})_{ij}%
\right] E^{j}+({\kappa}_{DB}) _{ij}B^{j}\,, \\[1ex]
H^{i}& =\left[{\mu }^{-1}{\delta }_{ij}+({\kappa}_{HB}) _{ij}%
\right] B^{j}+({\kappa }_{HE})_{ij}E^{j}\,.
\end{align}
\end{subequations}
These generalized scenarios lead to an unusual electrodynamics where the
electric displacement field receives a contribution from the magnetic flux density and the
magnetic field gets a contribution from the electric field. Such
modified constitutive relations, $\mathbf{{D}={D}}(\mathbf{{E},{B})}$ and $%
\mathbf{{H}={H}}(\mathbf{{E},{B})}$ are observed, for example, for
topological insulators \cite{Urrutia2}.

At this point, we introduce a generalized Ohm's law for the current
density $\mathbf{J}$, considering the contribution of the magnetic
current, $\mathbf{J}_{\mathrm{CME}}={\sigma}^{B}\cdot \mathbf{B}$, that is,
\begin{equation}
J^{i}=\sigma _{ij}E^{j}+{\sigma}_{ij}^{B}B^{j}\,,  \label{eq20}
\end{equation}%
composed of the usual Ohmic term involving the conductivity
tensor $\sigma_{ij}$ as well as an exotic term with the
magnetic conductivity tensor ${\sigma}_{ij}^{B}$, which has found realization in some
condensed matter systems. The second term can also be proposed as an extension of Ohm's law in
magnetohydrodynamics~\cite{Gurnett} (see also Sec.~\ref{section1}) as well as in
Lorentz-violating scenarios~\cite{Bailey}.

The magnetic conductivity tensor, ${\sigma}_{ij}^{B}$, is even under time reversal
($T$) and charge conjugation ($C$), but odd under parity transformations ($P$); see
Table.~\ref{tab:cpt-classification}.
Most notably, this tensor is even under time reversal, since
$\mathbf{J}$ and $\mathbf{B}$ are $T$ odd, which is highly unusual for a
conductivity. This behavior is analog to that of the phenomenological
parameter $\mu$ (not to be confused with the magnetic permeability)
observed in London's superconductivity model ($\mathbf{J}%
=-{\mu }^{2}\mathbf{A}$) \cite{Kharzeev2}. The $T$-even character of ${%
\sigma }_{ij}^{B}$ is typical of nondissipative and reversible processes~\cite{Kharzeev1, Kharzeev2}.
This property is what distinguishes the
magnetic conductivity from the usual Ohmic conductivity, which is $T$ odd and compatible with dissipative and nonreversible phenomena.
 \begin{table}[h]
\caption{Behavior of the Ohmic and exotic conductivity,
respectively, under $C$, $P$, and $T$ transformations.}
\begin{centering}
\begin{tabular}{ C{1.25cm}  C{1.25cm}  C{1.25cm}  C{1.25cm}  C{1.25cm}  C{1.25cm}}
\toprule
& $\textbf{E}$ & $\textbf{B}$ & $\textbf{J}$ & ${\sigma}$  & ${\sigma}^{B}$  \\[0.6ex]
\colrule
$C$ & $-$          & $-$          & $-$          & $+$ & $+$  \\ [0.6ex]
$P$ & $-$          & $+$          & $-$          & $+$ & $-$ \\[0.6ex]
$T$ & $+$          & $-$          & $-$          & $-$ & $+$  \\ [0.6ex]
\botrule
\end{tabular}
\end{centering}
\label{tab:cpt-classification}
\end{table} Table~\ref{tab:cpt-classification} shows a comparative analysis between the magnetic and Ohmic conductivity tensors under discrete symmetry
transformations, revealing crucial differences when the former
is subject to $T$, $P$, $CP$, and $CT$ transformations.

Using the generalized Ohm's law of \eqref{eq20} and conventional isotropic
constitutive relations, $D^{i}=\epsilon {\delta }_{ij}E^{j}$,
$H^{i}=\mu^{-1}{\delta}_{ij}B^{j}$, in the Maxwell equations,
\eqref{eq10} keeps its general form:
\begin{subequations}
\begin{equation}
\left[ \mathbf{k}\times \mathbf{k}\times {\mathbf{E}}\right] ^{i}+{\omega }%
^{2}\mu \bar{\epsilon}_{ij}(\omega )E^{j}=0\,,  \label{eq21}
\end{equation}%
where
\begin{equation}
{\bar{\epsilon}}_{ij}(\omega )=\left( \epsilon +\mathrm{i}{\frac{\sigma }{\omega }}%
\right) {\delta }_{ij}+{\frac{\mathrm{i}}{{\omega }^{2}}}({\sigma }^{B})_{ia}{%
\epsilon }_{abj}k_{b}\,,  \label{eq22}
\end{equation}%
\end{subequations}
defines the frequency-dependent extended permittivity tensor (EPT).
Equation~(\ref{eq21}) implies
\begin{equation}
\lbrack \mathbf{k}^{2}{\delta }_{ij}-k_{i}k_{j}-{\omega }^{2}\mu {\bar{{%
\epsilon }}}_{ij}]E^{j}=0\,.  \label{eq29}
\end{equation}%
Notice that the latter equation yields $k_i\bar{\epsilon}_{ij}E^{j} =0$, where we can
interpret $\bar{D}^i=\bar{\epsilon}_{ij}E^j$ as an extended displacement vector.

For a general anisotropic continuous scenario, we write
$\mathbf{k}=\omega \mathbf{n}$ where $\mathbf{n}$ is a vector pointing along
the direction of the wave vector and yielding the refractive index:\footnote{Here we again
take into account that the norm $|\mathbf{n}|$ is non-negative. To permit complex refractive
indices, we consider $+\sqrt{\mathbf{n}^2}$ instead of $|\mathbf{n}|$. The plus sign
indicates that we discard refractive indices with negative real parts.}
$n=+\sqrt{\mathbf{n}^2}$. Hence, Eq.~(\ref{eq29}) becomes
\begin{equation}
\left[n^2{\delta }_{ij}-n^{i}n^{j}-\mu \bar{{\epsilon }}_{ij}%
\right] E^{j}=0.  \label{eq30}
\end{equation}%
The latter can also be cast into the form
\begin{subequations}
\label{MijE-whole}
\begin{equation}
M_{ij}E^{j}=0\,,  \label{MijE}
\end{equation}%
where the tensor $M_{ij}$ reads
\begin{equation}
M_{ij}=n^{2}{\delta }_{ij}-n_{i}n_{j}-\mu \bar{{\epsilon }}_{ij}\,, \label{eq36B}
\end{equation}%
\end{subequations}
and $\bar{{\epsilon }}_{ij}$ is given by \eqref{eq22}. The set of
equations given above has nontrivial solutions for the electric field
if the determinant of the coefficient matrix $M_{ij}$ vanishes.
This condition on the determinant provides the dispersion relations associated
with wave propagation in the medium.

In the following, we will discard refractive indices with negative real parts
associated with frequencies that have the same property. Sophisticated composites of
different materials can be designed that have negative permittivity and permeability.
These are called metamaterials \cite{Kshetrimayum} and the real parts of their refractive indices must be
endowed with $\pm$ signs according to the materials at the interface:
material-material (+), material-metamaterial ($-$), or metamaterial-metamaterial (+).
On the contrary, negative refractive indices are not known to occur in crystals found
in nature, which is our focus in this paper.

\section{\label{section3}Propagation behavior under chiral and exotic
magnetic conductivity}

In this section, we will investigate the effects stemming from chiral
as well as exotic magnetic conductivities, incorporated into the formalism used for
describing the propagation of electromagnetic waves in
anisotropic dispersive media. In
order to examine the magnetic current $J_{\mathrm{CME}}^{i}={%
\sigma }_{ij}^{B}B^{j}$ classically, we consider the conductivity tensor ${\sigma}%
_{ij}^{B}$ in Eqs.~(\ref{eq22}) and (\ref{MijE-whole}) originating from the emergence of a magnetic field.
Studies of the CME \cite{Kharzeev1,Fukushima,Kharzeev2,Qiu} have reported the
generation of an electric current induced by a magnetic field,
\begin{equation}
J_{\mathrm{CME}}^{i}={\frac{e^{2}}{{4{\pi }^{2}}}}({\Delta }\mu )B^{i}\equiv \Sigma B^i\,,  \label{eq70}
\end{equation}%
where $e$ is the fermion charge, $\mathbf{B}$ the applied magnetic flux density,
${\Delta }\mu \equiv {\mu }_{R}-\mu _{L}$ is also known as the chiral
chemical potential, and $\Sigma$ is the chiral magnetic conductivity~\cite{Maxim,Leite,Dvornikov,Bubnov,Akamatsu,Maxim1,Maxim2,Burkov}.
As written in \eqref{eq70}, this effect is clearly represented by an isotropic diagonal
magnetic conductivity, that is,
\begin{subequations}
\begin{equation}
{\sigma}_{ij}^{B}={\Sigma \delta }_{ij}\,, \label{eq72}
\end{equation}%
in which
\begin{equation}
{\Sigma}={\frac{e^{2}}{{4{\pi }^{2}}}}{\Delta }\mu\,.  \label{eq74}
\end{equation}%
\end{subequations}
We can write down the magnetic conductivity ${\sigma }_{ij}^{B}$ in
the following form
\begin{equation}
{\sigma}_{ij}^{B}={\Sigma \delta}_{ij}+{\Sigma}_{ij}\,,  \label{eq69}
\end{equation}%
where ${\Sigma }$ is $1/3$ of the trace of the ${\sigma}_{ij}^{B}$ matrix and
represents the isotropic part of this conductivity, while ${\Sigma }_{ij}$
stands for all off-diagonal components of ${\sigma }_{ij}^{B}$. Hence, the
diagonal piece of the conductivity tensor is related to the CME and
constitutes the first case to be analyzed. We further examine the generalization of
the magnetic conductivity to exotic scenarios,
\begin{equation}
J^{i}={\Sigma}_{ij}B^{j},  \label{eq75}
\end{equation}%
where ${\Sigma}_{ij}$ comprises off-diagonal or anisotropic conductivity
components. It is worthwhile to note that the anisotropic chiral magnetic effect
\cite{Wurff} represents an interesting theoretical possibility to be
proposed and investigated. At first, it can be induced by a diagonal
anisotropic conductivity tensor, as will be examined in Sec.~\ref{diagonal-anisotropic}.

\subsection{\label{particular2}Isotropic diagonal chiral conductivity}

First of all, we discuss the behavior of an isotropic magnetic conductivity,
which is represented by a diagonal matrix; cf.~\eqref{eq72}. By inserting
the latter into \eqref{eq22}, one obtains
\begin{equation}
{\bar{\epsilon}}_{ij}(\omega )=\left( \epsilon +\mathrm{i}{\frac{\sigma }{\omega }}%
\right) {\delta }_{ij}-{\frac{{\mathrm{i}\Sigma }}{{\omega }^{2}}}{\epsilon }%
_{ijb}k_{b}\,,  \label{eq61a}
\end{equation}%
where the last term of \eqref{eq61a} represents the contribution from the
chiral conductivity. Note that we are starting from an isotropic permittivity
tensor, $\epsilon {\delta}_{ij}$, where all effects that are usually related to anisotropies in media manifest
themselves via the way the magnetic conductivity is coupled to the fields.
In this case, the tensor $M_{ij}$ given by \eqref{eq36B}, has the
form
\begin{widetext}
\begin{equation}
\label{eq62}
[M_{ij}]=\left(\begin{array}{ccc}
n_{2}^{2}+n_{3}^{2}-\mu\epsilon-\mathrm{i}\mu\frac{\sigma}{\omega} & -n_{1}n_{2}+\mathrm{i}\mu\frac{n_{3}\Sigma}{\omega} & -n_{1}n_{3}-\mathrm{i}\mu\frac{n_{2}\Sigma}{\omega}\\
\\
-n_{1}n_{2}-\mathrm{i}\mu\frac{n_{3}\Sigma}{\omega} & n_{1}^{2}+n_{3}^{2}-\mu\epsilon-\mathrm{i}\mu\frac{\sigma}{\omega} & -n_{2}n_{3}+\mathrm{i}\mu\frac{n_{1}\Sigma}{\omega}\\
\\
-n_{1}n_{3}+\mathrm{i}\mu\frac{n_{2}\Sigma}{\omega} & -n_{2}n_{3}-\mathrm{i}\mu\frac{n_{1}\Sigma}{\omega} & n_{1}^{2}+n_{2}^{2}-\mu\epsilon-\mathrm{i}\mu\frac{\sigma}{\omega}
\end{array}\right).
\end{equation}
\end{widetext}
Requiring $\det[M_{ij}]=0$, we get
\begin{subequations}
\begin{align}
\label{eq63}
n_{\pm}^{2}&=4\Upsilon_{\Sigma}+\mu\left(
-\epsilon+\mathrm{i}{\frac{\sigma}{\omega}}\right) \pm \frac{\mu\Sigma}{\omega}
\sqrt{2\Upsilon_{\Sigma}+\mathrm{i}\mu\frac{\sigma}{\omega}}\,, \\
\label{eq:quantity-upsilon}
2\Upsilon_{\Sigma}&=\mu\epsilon+\left(\frac{\mu}{2\omega}\Sigma\right)^2\,,
\end{align}
\end{subequations}
which can be split into real and imaginary parts as follows:
\begin{subequations}
\begin{equation}
n_{\pm}^{2}=\mu \epsilon+\frac{\mu\Sigma}{{\omega}}\left(\frac{\mu\Sigma}{2\omega}\pm N_+\right) +\mathrm{i}\frac{\mu}{\omega}(\sigma\pm\Sigma N_-)\,,  \label{eq63b}
\end{equation}
where
\begin{equation}
\label{63c}
N_{\pm}=\sqrt{\sqrt{\Upsilon_{\Sigma}^2+\left(\frac{\mu\sigma}{2\omega}\right)^2}\pm \Upsilon_{\Sigma}}\,.
\end{equation}
\end{subequations}
Equation~(\ref{eq63b}) yields two distinct refractive indices for each
frequency $\omega$, which is compatible with the physics of a conducting dielectric
medium endowed with birefringence. This behavior already occurs for the isotropic
conductivity tensor of~\eqref{eq72}, revealing that birefringence comes from
the way the chiral conductivity is coupled to the fields. Thus, when
considered within a medium of Ohmic conductivity ($\sigma \neq 0)$, a chiral
conductivity modifies the refractive index of the conducting medium.
It alters the phase velocity associated with its real part and the
absorption (or attenuation) coefficient related to its imaginary part.

\subsubsection{Dielectric nonconducting medium}

In the case  where we start from a
dielectric medium with zero Ohmic conductivity, $\sigma =0$, \eqref{eq63}
provides two distinct real values for the refractive index,
\begin{equation}
n_{\pm}^{2}=4\Upsilon_{\Sigma}-\mu\epsilon\pm %
\frac{\mu\Sigma}{\omega}\sqrt{2\Upsilon_{\Sigma}}\,,
\label{eq64B}
\end{equation}
which characterizes a dispersive nonconducting behavior typical of a
uniaxial crystal, where the propagation occurs with different velocities along
the principal dielectric directions \cite{Landau}. Therefore, the system will
behave like a birefringent dispersive dielectric medium where
electromagnetic waves propagate without undergoing attenuation (nonconducting
or absorbing behavior). The corresponding refractive indices
are given by
\begin{equation}
n_{\pm}=\sqrt{2\Upsilon_{\Sigma}}\pm \frac{\mu\Sigma}{2\omega}\,,
\label{n23-1}
\end{equation}
with $\Upsilon_{\Sigma}$ of \eqref{eq:quantity-upsilon}. The square of the latter leads
back to \eqref{eq64B}. In the present configuration, it is important to point
out that the chiral conductivity implies a typical conducting behavior for the
medium only when it is defined simultaneously with the Ohmic conductivity $(\sigma \neq
0,\sigma ^{B}\neq 0) $, as shown in the complex refractive index of \eqref{eq63b}.
When it is defined for a nonconducting dielectric $(\sigma =0,\sigma ^{B}\neq 0)$,
the behavior remains that of a dispersive nonabsorbing medium, as indicated by the
complex refractive index in \eqref{eq64B}. This is because $N_-=0$ when $\sigma=0$.

Alternatively, a refractive index can be determined from the frequency as a function
of the wave vector, $\omega=\omega(\mathbf{k})$, via the definition
$n\equiv +\sqrt{\mathbf{k}^2}/\omega(\mathbf{k})$ [see \eqref{eq13}]. The possible frequencies $\omega$
also follow from the requirement that $\det[M_{ij}]=0$ and are associated with particular
modes of the electric field. In order to better examine the features of propagation, we
implement $n=\sqrt{\mathbf{k}^2}/\omega$ in \eqref{eq64B} and obtain the dispersion equation
\begin{equation}
{\omega }^{4}-2{\omega }^{2}{\frac{{k^{2}}}{{\mu \epsilon}}}+\left( {\frac{{%
k^{2}}}{{\mu \epsilon }}}-\frac{\mu\Sigma^2}{2\epsilon}%
\right) ^{2}-{\frac{{{\mu}^{2}{\Sigma }^{4}}}{{4{\epsilon }^{2}}}}=0\,,
\label{eq64D}
\end{equation}
with $k\equiv \sqrt{\mathbf{k}^2}$. Solving for $\omega$, one gets
\begin{equation}
\omega^{2}_{\pm}={\frac{k^{2}}{{\mu \epsilon }}}\left( 1\pm {\frac{{\mu\Sigma}%
}{k}}\right)\,.  \label{eq64E}
\end{equation}
Equation~(\ref{eq64E}) represents two distinct modes, $\omega_{+}$ and $\omega
_{-}$. While the frequency $\omega _{+}$ is real for any value of $k$, the frequency
$\omega_{-}$ of the second mode can be imaginary if $k<\Sigma \mu $. To ensure that
$\omega _{-}$ represents the frequency of a physical propagating
mode, we should require that $k>\Sigma \mu $. Birefringence occurs when distinct
polarization modes propagate with different phase velocities. In this medium of zero
Ohmic conductivity, the phase velocities are
\begin{equation}
v_{\mathrm{ph}(\pm )}={\frac{{{\omega }_{\pm }}}{k}}={\frac{1}{\sqrt{\mu \epsilon}}}%
\sqrt{1\pm {\frac{{\mu\Sigma}}{k}}}\,,  \label{eq64F}
\end{equation}%
yielding the following phase velocity difference:
\begin{align}
\Delta v_{\mathrm{ph}}&={\frac{1}{\sqrt{\mu \epsilon }}}\left[ \sqrt{1+{\frac{{\mu\Sigma
}}{k}}}-\sqrt{1-{\frac{{\mu\Sigma}}{k}}}\,\right] \notag \\
& \simeq \frac{1}{\sqrt{%
\mu \epsilon }}{\frac{{\mu\Sigma}}{k}}\,,  \label{eq64G}
\end{align}
showing that the trace $\Sigma $ of the isotropic chiral conductivity is
really responsible for birefringence. Thus, a diagonal isotropic ${%
\sigma }_{ij}^{B}$ generates a birefringent dispersive nonconducting
behavior.

We can also analyze the effect that such a term has on the group velocity,%
\begin{equation}
v_{g(\pm)}=\left\vert {\frac{{\partial \omega_{\pm}}}{{\partial }\mathbf{k}}}%
\right\vert =\frac{1}{\sqrt{\mu \epsilon }}\frac{1\pm \mu\Sigma/(2k)}{\sqrt{1\pm \mu\Sigma/k}}\,,
\label{eq64H}
\end{equation}
which has a singularity for small momenta indicating problems with classical
causality due to $v_{g(\pm)}>1$. Causality is preserved for wave propagation
in the large-momentum regime.

One can also obtain the refractive indices for this medium ($\sigma =0)$ by
diagonalizing the permittivity tensor of \eqref{eq61a} and setting each
eigenvalue equal to $n^{2}/{\mu }$. In general, this procedure provides the refractive indices only
for the propagation along the directions of the principal axes of $\bar{\epsilon}_{ij}$.

The eigenvalues $\epsilon_{a}$ ($a=1,2,3$), where $\bar{\epsilon}_{ij}e_a^j=\epsilon_a e_a^i$
(with eigenvectors $\mathbf{e}_a$), are given by
\begin{subequations}
\begin{align}
\epsilon_{1}& =\epsilon\,,  \label{eigenvalue1} \\
\epsilon_{2,3}&\equiv\epsilon_{\pm} =\epsilon \pm\frac{{\Sigma}}{\omega}n\,,
\label{eigenvalue2}
\end{align}%
\end{subequations}
which can be associated with the following refractive indices:
\begin{subequations}
\begin{align}
n^{2}&=\mu \epsilon\,,  \label{n1} \\
n_{\pm}^{2}&=\mu \epsilon \pm \frac{\mu\Sigma}{\omega}n_{\pm}\,.
\label{n23}
\end{align}%
\end{subequations}
Surprisingly, the latter result reproduces \eqref{n23-1}, stemming from $\det[M_{ij}]=0$ and valid for an
arbitrary direction, which means that the eigenvalues $\epsilon_{2}$ and $\epsilon_{3}$ correspond to
the refractive indices of the medium $n_+$ and $n_-$, respectively.

In the following, we explain this behavior. The proposed method of finding the refractive indices $n$ through
the equation $n^2=\mu \epsilon_a(n)$, where $\epsilon_a$ are the eigenvalues of the EPT ${\bar \epsilon}_{ij}$,
only works under the requirement described below. For a general vector $\mathbf{n}$, the related electric field ${\mathbf E}_a$,
which satisfies the condition $M_{ij} E_a^j=0$ according to \eqref{eq30}, must be such that ${\mathbf n}\cdot {\mathbf E}_a=0$.
In this case, diagonalizing $M$ is equivalent to diagonalizing ${\bar\epsilon}$ and the result $\mathbf{E}_a \sim \mathbf{e}_a $
follows. Notice that we must also have ${\mathbf n}\cdot {\mathbf e}_a=0 $. This situation is clearly illustrated
in the present case, where Eqs.~(\ref{eq30}) and (\ref{eq61a}) yield the general condition $\mathbf{n}\cdot \mathbf{E}=0$.
Here, the three eigenvectors of the generalized permittivity are
\begin{subequations}
\begin{align}
\label{eigenvector1}
\mathbf{e}_1&=\frac{\mathbf{n}}{n}\equiv\mathbf{m}\,, \\[2ex]
\label{eigenvector2}
\mathbf{e}_{2,3}&=\frac{1}{\sqrt{2(m_1^2+m_3^2)}}\begin{pmatrix}
m_3 \mp \mathrm{i}m_1 m_2 \\
\pm\mathrm{i}(m_1^2+m_3^2) \\
\mp\mathrm{i}m_2 m_3-m_1 \\
\end{pmatrix}\,,
\end{align}
\end{subequations}
with the unit vector $\mathbf{m}$ defining the planes of constant phase of the wave. Let us observe that the eigenvectors in
Eqs.~(\ref{eigenvector1}) and (\ref{eigenvector2}) are independent of the corresponding refractive indices $n_{\pm}$,
being just functions of the direction given by $\mathbf{m}$. We note that $\mathbf{e}_{1}\cdot \mathbf{e}_{2}^{*}=\mathbf{e}_{1}\cdot \mathbf{%
e}_{3}^{*}=\mathbf{e}_{2}\cdot \mathbf{e}_{3}^{*}=0$, whereupon these three eigenvectors are linearly independent.
In particular, ${\mathbf e}_2$ and ${\mathbf e}_3$ are orthogonal to ${\mathbf e}_1={\mathbf m}$, thus yielding
the correct refractive indices $n_{\pm}$ of \eqref{eq64B}, according to the proposed method. In this case, the
propagating modes of the electric field are correctly described by the eigenvectors ${\mathbf e}_2$ and ${\mathbf e}_3$
and the eigenvalue $\epsilon_1=\epsilon$ has to be rejected, because ${\mathbf m}\cdot {\mathbf e}_1$ is nonzero.

To decide which refractive indices are physical, we can also look at the modes of the electric field. The latter are
obtained from solving the homogeneous system of equations $M_{ij}E^j=0$ for $\mathbf{E}$ with $\omega$ replaced by
the dispersion relations $\omega(\mathbf{k})$ determined from the coefficient determinant.
In the dielectric nonconducting medium under consideration, a particular frequency does not correspond to a physical mode
when the electric field is longitudinal. In other words, a mode is unphysical when its electric field points along the direction
of the wave vector, i.e., $\mathbf{E}\cdot\mathbf{k}=|\mathbf{E}||\mathbf{k}|$ or
$\mathbf{k}\times\mathbf{E}=\mathbf{0}$. As $[M,\overline{\epsilon}]=0$ for the
particular isotropic configuration of \eqref{eq72}, the eigenvector of \eqref{eigenvector1}
also corresponds to a mode of the electric field. As it is longitudinal,
the associated solution for the permittivity $\epsilon_1=\epsilon$ cannot be
physical and must be discarded.

Since the configuration studied is isotropic, we can choose
$\mathbf{m}=(0,0,m)$ without loss of generality.
Equation~(\ref{eigenvector2}) then results in
\begin{equation}
\label{eq:polarizations-isotropic}
\mathbf{e}_{2,3}=\frac{1}{\sqrt{2}}\begin{pmatrix}
1 \\
\pm\mathrm{i} \\
0 \\
\end{pmatrix}\,.
\end{equation}
The vectors $\mathbf{e}_{2}$ and $\mathbf{e}_{3}$, respectively, can be interpreted as the polarization
vectors of left-handed ($L$) and right-handed ($R$) polarized electromagnetic waves.\footnote{We define
a polarization as right-handed (left-handed) if the
polarization vector of a plane wave rotates along a circle in the clockwise (counterclockwise)
direction when the observer is facing into the incoming wave \cite{Jackson, Zangwill}.}
Hence, we identify $n_{L,R}\equiv n_{\pm}$.

These polarizations are transverse, i.e., perpendicular
to $\mathbf{m}$. According to Table.~\ref{tab:cpt-classification}, the magnetic
conductivity $\sigma^B_{ij}$ is odd under parity transformations. While parity violation does
not show up in a single refractive index of \eqref{n23-1} on its own, it becomes manifest in the distinct
propagation properties of left- and right-handed polarized electromagnetic waves. This
is the physical reason for birefringence.

The behavior found for this configuration is highly interesting. Parity violation could
be expected to imply refractive indices that are angular dependent. In other words,
birefringence in a material is usually caused by the presence of at least a single
optical axis. An optical axis indicates a preferred direction in the crystal, whereupon
its refractive index cannot be isotropic, anymore. Hence, under usual circumstances,
an occurrence of parity violation and birefringence seems to contradict isotropy of a
crystal. If a magnetic conductivity is present, we found that birefringence can
also emerge in an isotropic crystal.

If birefringence occurs in a medium with a single
optical axis, a light ray can split into an ordinary and an extraordinary ray. The
ordinary one behaves according to Snell's law whereas the extraordinary one does not do
so. The electric field associated with the extraordinary ray is no longer orthogonal
to its wave vector. However, the polarizations of the electric field stated in
\eqref{eq:polarizations-isotropic} are orthogonal to the wave vector, which is why an
extraordinary ray cannot be identified in this setting. The two polarizations only split
due to their distinct propagation velocities, cf.~\eqref{eq64H}. Therefore, the polarization
plane of a linearly polarized wave will rotate indicating an optically active material.
This phenomena is quantified by defining the specific rotatory power $\Delta$, measuring
the rotation of the plane of linearly polarized light per unit traversed length in the
medium. It is given by $\Delta\equiv -(\Delta n)\omega/{2}$ where
$\Delta n\equiv n_{+}-n_{-}$ is the difference of refractive indices for the two
polarization directions. In our case, the latter definition yields
\begin{equation}
\label{eq:rotatory-power}
\Delta=-\frac{\mu\Sigma}{2}\,,
\end{equation}
thus providing a frequency-independent specific rotatory power due to the chiral magnetic
conductivity $\Sigma$.

As we know, there is no clear connection between the CME and measurements
of birefringence, and so we discuss a possible way to observe this effect. Before
more elaborate methods were developed, birefringence was detected
with a set of crossed polarizers that does not permit any light to pass. If a birefringent
material is put in between these two polarizers, it rotates the polarization plane of
the light that passed the first polarizer such that light of a nonvanishing intensity
can be measured behind the second polarizer. However, this method is not precise enough to
measure small values of birefringence. Furthermore, it is challenging to measure birefringence
of an inhomogeneous medium.

Therefore, a more sophisticated technique was developed in Ref.~\cite{Birefringence1}.
The system to be used is based on a rotating polarizer, a quarter-wave plate, and an analyzer.
The passing light is captured by a CCD camera whereupon the measured data are evaluated by a
computer.  This system was commercially distributed under the name \textit{Metripol}.

The intensity $I$ of the light detected at any point is expressed in terms of angles
$\xi$ and $\psi$ describing the orientation of the rotating polarizer:
\begin{equation}
I=\frac{I_0}{2}[1-\sin(2\xi-2\psi)\sin\delta]\,,
\end{equation}
where $I_0/2$ is a suitably normalized intensity and $\delta$ is the phase shift between
the two physical polarizations of the light. The latter is
\begin{equation}
\delta=\frac{2\pi}{\lambda}d(\Delta n),
\label{phase-shift2}
\end{equation}%
where $d$ is the sample thickness and $\lambda$ is the vacuum wavelength of the incoming light.

Thus, to measure birefringence of a material endowed with a magnetic
conductivity, a sample of thickness $d$ can be placed into a \textit{Metripol} system to determine
the factor $\sin\delta$. This factor provides the difference between the refractive indices via
$\Delta n=\delta\lambda/(2\pi d)$. The latter experimental result in combination with Eq.~(\ref{n23-1})
leads to the magnetic-conductivity parameter:
\begin{equation}
\Sigma=\frac{\omega(\Delta n)}{\mu}\,.
\end{equation}
A similar technique for birefringence measurements was also reported in Ref.~\cite{Birefringence2}.

\vspace{-0.5cm}\subsection{\label{diagonal-anisotropic}Diagonal anisotropic chiral
conductivity}

As a next step, we will consider another particular case for the chiral
conductivity, ${\sigma }_{ij}^{B}$, which is represented by a diagonal
tensor that describes an anisotropic system \cite{Chang}:
\begin{equation}
\left[ {\sigma }_{ij}^{B}\right] =%
\begin{pmatrix}
{\Sigma }_{x} & 0 & 0 \\
0 & {\Sigma }_{y} & 0 \\
0 & 0 & {\Sigma }_{z}%
\end{pmatrix}\,,  \label{d1}
\end{equation}%
with a set $\{\Sigma\}=\{\Sigma_x,\Sigma_y,\Sigma_z\}$ of distinct
elements in the diagonal, ${\Sigma }_{x}\neq {\Sigma }%
_{y}\neq {\Sigma }_{z}$. The general
permittivity tensor is given by \eqref{eq22}, with the components shown
as follows:
\begin{equation}
\left[ \bar{\epsilon}_{ij}\right] =%
\begin{pmatrix}
\epsilon +\mathrm{i}{\frac{\sigma }{\omega }} & -{\frac{\mathrm{i}}{{\omega }^{2}}}{\Sigma }%
_{x}k_{3} & {\frac{\mathrm{i}}{{\omega }^{2}}}{\Sigma }_{x}k_{2} \\
&  &  \\
{\frac{\mathrm{i}}{{\omega }^{2}}}{\Sigma }_{y}k_{3} & \epsilon +\mathrm{i}{\frac{\sigma }{%
\omega }} & -{\frac{\mathrm{i}}{{\omega }^{2}}}{\Sigma }_{y}k_{1} \\
&  &  \\
-{\frac{\mathrm{i}}{{\omega }^{2}}}{\Sigma }_{z}k_{2} & {\frac{\mathrm{i}}{{\omega }^{2}}}{%
\Sigma }_{z}k_{1} & \epsilon +\mathrm{i}{\frac{\sigma }{\omega }}%
\end{pmatrix}\,.  \label{d6}
\end{equation}
The dispersion relations are obtained from $\det[M_{ij}]=0$ where the tensor
$M_{ij}$ is defined in \eqref{eq36B}, with $\bar{\epsilon}_{ij}$ given
by \eqref{d6}.
Evaluating this condition leads to the dispersion equation
\begin{subequations}
\begin{equation}
\left[\omega n^{2}-\mu(\mathrm{i}\sigma +\epsilon\omega)\right]^{2}=\Omega\,,
\label{d10}
\end{equation}
with the function
\begin{equation}
\label{eq:function-Omega}
\Omega=\Omega(\mathbf{n})={\mu }^{2}({\Sigma }_{x}{\Sigma }_{y}n_{3}^{2}+{\Sigma }_{x}{\Sigma }_{z}n_{2}^{2}+{\Sigma }_{y}{\Sigma }_{z}n_{1}^{2})\,.
\end{equation}
\end{subequations}
The latter is an involved expression, since it explicitly contains the components
of the vector $\mathbf{n}=(n_{1},n_{2},n_{3})$ instead of its modulus $n$ as in the
isotropic case investigated previously.
To avoid problems with separating real and imaginary parts of refractive
indices, we will assume that $\Omega\geq 0$. The following parametrization~\cite{Zangwill}
for $\mathbf{n}$ permits a convenient examination of the physical content of \eqref{d10}:
\begin{equation}
\mathbf{n}=n(\sin \theta \cos \phi ,\sin \theta \sin \phi ,\cos \theta)\equiv n \, {\mathbf m}\,,  \label{d15}
\end{equation}%
with angles $\theta\in [0,\pi]$ and $\phi\in [0,2\pi)$.
With the latter, \eqref{d10} is rewritten as
\begin{subequations}
\label{d17a}
\begin{equation}
{\omega }^{2}n^{4}-Dn^{2}-G=0\,,
\end{equation}%
where
\begin{align}
D&=2\mu\omega(\epsilon\omega+\mathrm{i}\sigma)+\tilde{\Omega}\,,  \label{d18} \\
G&=\mu^2(\sigma^2-2\mathrm{i}\epsilon\omega\sigma-\epsilon^2\omega^2)\,,  \label{d19} \\
\label{eq:definition-tilde-Omega}
\tilde{\Omega}&=\mu ^2\left[\cos^2\theta \Sigma _x\Sigma _y+
\sin^2\theta\sin^2\phi \Sigma _x\Sigma_z\right.  \notag \\
&\phantom= \left.+\sin^2\theta\cos^2\phi \Sigma _y\Sigma _z\right]=\Omega/n^2\,.
\end{align}
\end{subequations}
The dispersion equation~(\ref{d17a}) yields two solutions for $n$,
\begin{equation}
n^{2}_{\pm}={\frac{D}{{2{\omega }^{2}}}}\pm {\frac{1}{{2{\omega }^{2}}}}\sqrt{{%
D^{2}}+4{\omega }^{2}G}\,,  \label{d20}
\end{equation}%
revealing two values for any frequency.

\subsubsection{Dielectric nonconducting medium}

In the special limit of zero Ohmic
conductivity, $\sigma =0$, all the complex pieces of \eqref{d10}
or Eq.~(\ref{d17a}) vanish and the refractive indices of \eqref{d20} are real.
Some complex term could occur in \eqref{d20} if the discriminant is negative: $\Delta _{0}=D_{0}^{2}+4%
{\omega }^{2}G_{0}<0,$ where $G_{0}=-{\epsilon }^{2}{\mu }^{2}{\omega }^{2}$
and $D_{0}=2\epsilon \mu {\omega }^{2}+\tilde{\Omega} .$ But $\Delta _{0}={\tilde{\Omega}}
^{2}+4\tilde{\Omega} \epsilon \mu {\omega }^{2}>0$ and therefore, this possibility is not
realized. In fact, for $\sigma =0$, Eq.~(\ref{d10}) gives rise to
\begin{equation}
\omega^2(n^{2}-\mu\epsilon)^{2}=\Omega\,,
\label{d30}
\end{equation}%
whereupon
\begin{equation}
n^{2}_{\pm}=\mu\epsilon\pm\frac{1}{{{\omega }}}\sqrt{\Omega}\,.
\label{d31}
\end{equation}
If we apply the prescription stated in \eqref{d15} to \eqref{d20}, the refractive indices
are given by
\begin{equation}
n^{2}_{\pm}=\mu\epsilon+\frac{\tilde{\Omega}}{2\omega^2}\pm \frac{1}{2 \omega^2}\sqrt{4\mu\epsilon \omega^2{\tilde \Omega}+{\tilde{\Omega}}^2}\,.
\label{d31b}
\end{equation}
It is interesting to observe that by using the parametrization of \eqref{d15} in each refractive
index of \eqref{d31} one obtains an explicit solution for $n_{\pm}$. Each equation is of
second order and the refractive indices are
\begin{subequations}
\label{eq:refractive-index-anisotropic-diagonal-case}
\begin{align}
n_{\pm}&=\sqrt{2\Upsilon_{\{\Sigma\}}}\pm\frac{\sqrt{\tilde\Omega}}{2\omega}\,, \displaybreak[0]\\[2ex]
\label{eq:quantity-upsilon-anisotropic}
2\Upsilon_{\{\Sigma\}}&=\mu\epsilon+\frac{\tilde\Omega}{4\omega^2}\,,
\end{align}
\end{subequations}
where one easily verifies that their squares reproduce the results in \eqref{d31b}.
Hence, for $\sigma=0$, $\sigma^{B}_{ij}\neq 0$, we have a
dielectric nonconducting and dispersive medium. One can verify that
\eqref{d30} provides three different expressions for the following three
situations: $\mathbf{n}=(0,n_{2},n_{3}),$ $\mathbf{n}=(n_{1},0,n_{3})$,
and $\mathbf{n}=(n_{1},n_{2},0)$.
For each of these three choices, \eqref{d31} yields two values for $n$,
resulting in birefringence. 

Diagonalizing the EPT of~\eqref{d6} for zero Ohmic conductivity ($\sigma
=0)$ and dropping the parametrization given by \eqref{d15}, one obtains
$\epsilon _{1}=\epsilon$ and
\begin{subequations}
\begin{equation}
\epsilon_{2,3}=\epsilon\pm \frac{\gamma}{\omega^{2}}\,, \label{de1a}
\end{equation}
with
\begin{align}
\gamma(\mathbf{k})&\equiv\frac{\omega}{\mu}\sqrt{\Omega(\mathbf{k}/\omega)} \notag \\
&=\sqrt{\Sigma _{y}\Sigma _{z}k_{1}^{2}+\Sigma _{x}\Sigma_{z}k_{2}^{2}+\Sigma _{x}\Sigma _{y}k_{3}^{2}}\,.
\end{align}
\end{subequations}
In this case, when the two eigenvalues of the EPT, given in \eqref{de1a}, are multiplied
by $\mu$,  we obtain $\mu\epsilon_{2,3}=\mu \epsilon_{\pm}$, being equal to the values
for $n^2_{\pm}$ in \eqref{d31} calculated from the condition $\det[M_{ij}]=0$. This numerical
coincidence provides a nice example of the fact that even though we satisfy the condition
$n^2_{\pm}=\mu \epsilon_\pm $, this equality does not imply that the associated eigenvectors
of the EPT would necessarily correspond to the propagation modes of the electric field.
The corresponding eigenvectors are
\begin{subequations}
\begin{align}
\mathbf{e}_{1}&=\begin{pmatrix}
k_{1}/k_{3} \\
k_{2}/k_{3} \\
1 \\
\end{pmatrix}\,, \displaybreak[0]\\
\mathbf{e}_{2}&=\frac{1}{\varkappa}\begin{pmatrix}
\mathrm{i}\Sigma _{x}k_{2}\gamma -\Sigma_{x}\Sigma _{y}k_{1}k_{3} \\
-\mathrm{i}\Sigma _{y}k_{1}\gamma -\Sigma _{x}\Sigma_{y}k_{2}k_{3} \\
\varkappa
\end{pmatrix}\,, \label{eigen2} \displaybreak[0]\\
\mathbf{e}_{3}&=\frac{1}{\varkappa}\begin{pmatrix}
-\mathrm{i}\Sigma_{x}k_{2}\gamma -\Sigma_{x}\Sigma _{y}k_{1}k_{3} \\
\mathrm{i}\Sigma _{y}k_{1}\gamma -\Sigma _{x}\Sigma_{y}k_{2}k_{3} \\
\varkappa \\
\end{pmatrix}\,, \label{eigen3}
\end{align}
\end{subequations}
where
\begin{align}
\label{definition-varkappa}
\varkappa&=\Sigma_{z}(\Sigma_{y}k_{1}^{2}+\Sigma_{x}k_{2}^{2}) \notag \\
&=\omega^2 n^2\Sigma_{z}(\Sigma _{y}m_{1}^{2}+\Sigma _{x}m_{2}^{2})\,,
\end{align}
with $\mathbf{m}$ defined in \eqref{d15}. Note that
\begin{subequations}
\begin{align}
\label{product3}{\bf{e}}_{1}\cdot {\bf{e}}_{2}^{*}&=1-\frac{(k_{1}^{2}+k_{2}^{2}){\Sigma}_{x}{\Sigma}_{y}}{\varkappa} \notag \\
& \phantom{{}={}} -\mathrm{i}\frac{\gamma k_{1}k_{2}({\Sigma}_{x}-{\Sigma}_{y})}{\varkappa k_{3}}\,,\\
\label{product2}
{\bf{e}}_{1}\cdot {\bf{e}}_{3}^{*}& =\mathbf{e}_1^{*}\cdot \mathbf{e}_2\,, \\[2ex]
\label{product4} {\bf{e}}_{2}\cdot {\bf{e}}_{3}^{*}&=1-\frac{{\Sigma}_{x}^{2}(\gamma k_{2} +\mathrm{i}k_{1}k_{3}{\Sigma}_{y})^{2}}{{\varkappa}^{2}} \notag \\
&\phantom{{}={}}+\frac{{\Sigma}_{y}^{2}(k_{2}k_{3}{\Sigma}_{x}+\mathrm{i}\gamma k_{1})^{2}}{{\varkappa}^{2}}\,.
\end{align}
\end{subequations}
The above calculation shows that ${\mathbf k}\cdot {\mathbf e}_a \neq 0$ for all $a=1,2,3$
in such a way that these eigenvectors cannot describe the propagation modes ${\mathbf E}_{\pm}$
of the electric field. In order to clearly illustrate this point, we calculate such propagation
modes for the particular case where ${\mathbf k}=(k_1,k_2,0)=\omega \, n\,(m_1,m_2,0)$ and
compare those with the corresponding eigenvectors of the EPT. The results for the propagation
modes are
\begin{subequations}
\label{eq:electric-field-plus-minus}
\begin{align}
\mathbf{E}_{\pm}&=\begin{pmatrix}
-\omega n_{\pm}m_{2}/u_{\pm} \\
\omega n_{\pm}m_{1}/v_{\pm}  \\
\pm \mathrm i\omega^{2}\Sigma_{z}/\sqrt{\varkappa_{\pm}} \\
\end{pmatrix}\,,
\end{align}
with
\begin{align}
u_{\pm}&=\frac{\mu\epsilon}{1\mp \mu\Sigma_{x}\Sigma_{z}/\sqrt{\varkappa_{\pm}}}\,, \displaybreak[0]\\[2ex]
v_{\pm}&=\frac{\mu\epsilon}{1\mp \mu\Sigma_{y}\Sigma_{z}/\sqrt{\varkappa_{\pm}}}\,, \displaybreak[0]\\[2ex]
\varkappa_{\pm}&=\omega^2n_{\pm}^2\Sigma_z(\Sigma_ym_1^2+\Sigma_xm_2^2)\,.
\label{modeepm}
\end{align}
\end{subequations}
The refractive indices satisfy the quadratic equation
\begin{align}
n^2_\pm&=\mu\left(\epsilon\pm \frac{n_{\pm}}{\omega}\sqrt{\Sigma_z(\Sigma_x m_2^2+ \Sigma_y m_1^2)}\right)\,,\label{nm}
\end{align}
which is written in terms of the unit vector $\mathbf{m}$ defining the planes of constant phase of the wave.
The dispersion relations $\omega_{\pm}=\omega_{\pm}(\mathbf{k})$ are
\begin{align}
\omega_{\pm}&=\frac{1}{\sqrt{\mu\epsilon}}\sqrt{k_1^2+k_2^2\mp \mu\sqrt{\Sigma_z(\Sigma_x k_2^2+\Sigma_yk_1^2)}}\,.
\end{align}
The functions $u_{\pm}$ and $v_{\pm}$ satisfy the useful relations
\begin{equation}
n_{\pm}^2\left(\frac{m_1^2}{v_{\pm}}+\frac{m_2^2}{u_{\pm}}\right)=1\,,
\end{equation}
which are just a consequence of \eqref{nm}. We observe that
\begin{equation}
{\mathbf m}\cdot{\mathbf E}_{\pm}=\pm\epsilon^{-1}m_1m_2\frac{(\Sigma_x-\Sigma_y)\sqrt{\Sigma_z}}{\sqrt{\Sigma_ym_1^2+\Sigma_xm_2^2}}\,.
\end{equation}
Due to $\gamma|_{k_3=0}=\sqrt{\varkappa}$, from Eqs.~(\ref{eigen2}) and (\ref{eigen3}) we
read that in this case the eigenvectors of the EPT are
\begin{subequations}
\begin{align}
\mathbf{e}_{\pm}&\equiv\pm\mathrm{i}\sqrt{\varkappa}\mathbf{e}_{2,3}=\begin{pmatrix}
-\Sigma_{x}k_{2}  \\
\Sigma_{y}k_{1}  \\
\pm {\mathrm i}\sqrt{\varkappa} \\
\end{pmatrix}\,, \displaybreak[0]\label{eigen_EPT}\\[2ex]
\epsilon _{\pm}&=\epsilon \pm \frac{\sqrt{\varkappa}}{\omega^{2}}\,,
\end{align}
\end{subequations}
with $\varkappa$ of Eq.~(\ref{definition-varkappa}). The latter $\mathbf{e}_{\pm}$ do not coincide with
${\mathbf E}_\pm$ of \eqref{eq:electric-field-plus-minus}.
The propagating modes of \eqref{eq:electric-field-plus-minus} are not orthogonal to the
wave vector ${\mathbf k}$. This is a characteristic of wave propagation in anisotropic
media, which nevertheless must always fulfill ${\mathbf k}\cdot{\mathbf D}=0$ in the
absence of sources.

To have a direct physical interpretation of the propagating modes in
\eqref{eq:electric-field-plus-minus}, we perform another simplification and consider
the case of propagation along the $y$ axis, i.e., we take $\mathbf{m}=(0,1,0)$. Now we have
$n^2_{\pm}=u_{\pm}$ and $\sqrt{\varkappa_{\pm}}=\omega n_{\pm}\sqrt{\Sigma_x \Sigma_z}$.
After substituting in \eqref{eq:electric-field-plus-minus}, we obtain the normalized
electric fields
\begin{align}
\label{eq:eliptical-polarization}
\hat{\mathbf{E}}_{\pm}|_{m_1=m_3=0}&=\sqrt{\frac{\Sigma_x}{\Sigma_x+\Sigma_z}}\begin{pmatrix}
1 \\
0 \\
\mp\mathrm{i}\sqrt{\Sigma_z/\Sigma_x} \\
\end{pmatrix}\,,
\end{align}
which describe elliptically polarized modes with distinct refractive indices
\begin{equation}
n_{\pm}=\sqrt{\mu\epsilon+\frac{\mu^2}{4\omega^2}\Sigma_x\Sigma_z} \pm \frac{\mu}{2 \omega}\sqrt{\Sigma_x\Sigma_z}\,.
\label{EpmPart}
\end{equation}
The latter correspond to specific cases of \eqref{eq:refractive-index-anisotropic-diagonal-case}.
The modes associated with $\hat{{\bf{E}}}_{\pm}$ in \eqref{eq:eliptical-polarization} represent the left-handed and right-handed
polarization states via the same definition used at the end of Sec.~\ref{particular2}. It is relevant to
point out that the propagation of the wave associated with \eqref{eq:eliptical-polarization} occurs along the $y$ axis, while
the wave connected to \eqref{eq:polarizations-isotropic} propagates along the $z$ axis. This explains the opposite signs
that appear in the polarization modes of \eqref{eq:eliptical-polarization} in comparison with those of
\eqref{eq:polarizations-isotropic}.

Making cyclic changes among $\{x,y,z\}$, we identify $n_{\pm}=n_{L,R}$, thus characterizing an
optically active medium. Note that in this particular case we have $\mathbf{k} \cdot \hat{\mathbf{E}}_{\pm}=0$,
so we expect $\hat{\mathbf{E}}_{\pm}$ to be proportional to the eigenvectors $\mathbf{e}_{\pm}$ in \eqref{eigen_EPT},
as can be readily verified.

It is also easy to notice that Eq.~(\ref{eq:refractive-index-anisotropic-diagonal-case}) together with
\eqref{EpmPart} include the isotropic case. Indeed, assuming that ${\Sigma}_{x}={\Sigma}_{y}={\Sigma}_{z}={\Sigma}$,
we have $\tilde{\Omega}=\mu^2\Sigma^2$ and Eq.~(\ref{eq:refractive-index-anisotropic-diagonal-case}) exactly reproduces
\eqref{n23-1}. Also, we recover the circularly polarized modes of \eqref{eq:polarizations-isotropic},
though with respect to a different axis. Both the isotropic and anisotropic nonconducting
diagonal cases exhibit birefringence.

Still, a main difference arises: in the former case, the two refractive indices are independent of the direction
defined by the wave vector $\mathbf{k}$ [see \eqref{eq64B}], while in the latter case
a direction dependence emerges due to the appearance of $\tilde\Omega$ in \eqref{d31b}. Nevertheless, $\tilde\Omega$
is a quadratic function of the components of the wave vector $\mathbf{k}$, which yields a parity-invariant individual
refractive index. As in the isotropic case, parity violation manifests itself in the different refractive indices of
the left- and right-handed modes. Another remarkable feature of the diagonal anisotropic case, which
will also appear in the instances discussed below, is the presence of a non-Hermitian EPT.
Although the latter opens up interesting new possibilities, we defer related analyses to future work.

Finally, it is intriguing to notice that a diagonal chiral conductivity tensor inserted into the Maxwell equations
is not enough to establish a conducting behavior for a dielectric medium although this is the chiral conductivity
configuration most used and cited in the literature.\\

\subsection{Nondiagonal antisymmetric conductivity}
\label{sec:nondiagonal-antisymmetric-configuration}

Now we analyze the case where the magnetic conductivity is described by an exotic
configuration given by an antisymmetric tensor, ${\sigma }_{ij}^{B}$, parametrized
in terms of a three-vector, $\mathbf{b}=(b_{1},b_{2},b_{3})$, as
\begin{equation}
{\sigma}_{ij}^{B}={\epsilon}_{ijk}b_{k}\,,  \label{eq65}
\end{equation}
with the Levi-Civita symbol $\epsilon_{ijk}$ in three dimensions. Notice that this
case corresponds to the situation in magnetohydrodynamics of an isotropic resistivity
in \eqref{ROL} and the vector $\mathbf{b}$ is proportional to the plasma velocity.
Inserting \eqref{eq65} into \eqref{eq22}, leads to the EPT
\begin{equation}
{\bar{\epsilon}}_{ij}(\omega)=\left(\epsilon+\mathrm{i}{\frac{\sigma }{\omega }}+\mathrm{i}
\frac{{\mathbf{k}\cdot \mathbf{b}}}{{\omega }^{2}}\right) {\delta }_{ij}-{%
\frac{\mathrm{i}}{{\omega}^{2}}}k_{i}b_{j}\,.  \label{eq66}
\end{equation}%
By employing \eqref{eq66} in \eqref{eq36B}, one obtains
\begin{widetext}
\begin{align}
\label{eq66-1}
[M_{ij}]&=\begin{pmatrix}
n_2^2+n_3^2-\mu\epsilon & -n_1n_2 & -n_1n_3 \\
-n_1n_2 & n_1^2+n_3^2-\mu\epsilon & -n_2n_3 \\
-n_1n_3 & -n_2n_3 & n_1^2+n_2^2-\mu\epsilon \\
\end{pmatrix} \notag \\
&\phantom{{}={}}+\mathrm{i}\frac{\mu}{\omega}\begin{pmatrix}
-(\sigma+n_2b_2+n_3b_3) & n_1b_2 & n_1b_3 \\
n_2b_1 & -(\sigma+n_1b_1+n_3b_3) & n_2b_3 \\
n_3b_1 & n_3b_2 & -(\sigma+n_1b_1+n_2b_2) \\
\end{pmatrix}\,,
\end{align}
\end{widetext}
for which $\det[M_{ij}]=0$ implies
\begin{equation}
\left[n^{2}-\mathrm{i}{\frac{\mu }{\omega }}\left(\mathbf{b}\cdot\mathbf{n}\right)
-\mu \left(\epsilon+\mathrm{i}\frac{\sigma}{\omega}\right)\right]^{2}=0.
\label{eq67B}
\end{equation}
Contrary to Eqs.~(\ref{d17a}) and (\ref{d30}), the latter dispersion equation
involves the square of a quadratic polynomial in the components of $\mathbf{n}$. Hence, the
solution for the refractive index is doubly degenerate and there is only a single refractive
index with a non-negative real part.
Implementing $\mathbf{b}\cdot \mathbf{n}=bn\cos \theta $ with $b=|\mathbf{b}|$,
\eqref{eq67B} provides the following refractive index:
\begin{subequations}
\label{eq68}
\begin{align}
n&=\sqrt{2\Upsilon_b+\mathrm{i}\mu\frac{\sigma}{\omega}}+\mathrm{i}\Xi_b\,,
\label{eq67a} \displaybreak[0]\\
\label{gamma}
2\Upsilon_b&=\mu\epsilon-\Xi_b^2\,,\quad \Xi_b=\frac{\mu}{2\omega}b\cos\theta\,.
\end{align}
\end{subequations}
Note the presence of the minus sign between the two contributions in $\Upsilon_b$
in contrast to $\Upsilon_{\Sigma}$ of \eqref{eq:quantity-upsilon} and $\Upsilon_{\{\Sigma\}}$ in
\eqref{eq:quantity-upsilon-anisotropic}. Thus, we assume that $\mu\epsilon\geq \Xi_b^2$.
Decomposing the latter refractive index into its real and imaginary parts implies
\begin{subequations}
\begin{equation}
n=\alpha^{\prime}_++\mathrm{i}\left(\Xi_b+\alpha^{\prime}_-\right)\,,  \label{eq68-a}
\end{equation}
where
\begin{equation}
\alpha^{\prime}_{\pm}=\sqrt{\sqrt{\Upsilon_b^2+\left(\frac{\mu\sigma}{2\omega}\right)^{2}}\pm \Upsilon_b}\,.
\label{eq68-b}
\end{equation}
\end{subequations}
In this case, we obtain an anisotropic complex refractive index that
captures the effects of the exotic conductivity, as shown in
\eqref{eq68-a}, which is compatible with a conducting medium. The imaginary
part takes the role of the absorption coefficient for the electromagnetic wave,
which undergoes attenuation while it propagates. Therefore, an electromagnetic
wave cannot propagate through such a medium, since absorption
damps its intensity. When considered in a dielectric conducting
medium, $(\epsilon ,\sigma ,\sigma ^{B})$, the magnetic
conductivity modifies the real and imaginary parts of the refractive index.
This causes a modification of the absorption coefficient given by ${\tilde{%
\alpha}}={\omega }\alpha^{\prime}_-+\mu b\cos \theta$, where $\alpha^{\prime}_-$ is
stated in \eqref{eq68-b}.

Note that for a diagonal magnetic conductivity there were two distinct
refractive indices with a positive real part; see Eqs.~(\ref{n23-1}) and (\ref{eq:refractive-index-anisotropic-diagonal-case}).
Thus, the occurrence of a single refractive index in Eq.~(\ref{eq68}) is quite unexpected in
the context of a parity-violating theory. In contrast to the cases studied
before, a magnetic conductivity given by \eqref{eq65} does not imply birefringence.
These results suggest that an antisymmetric magnetic conductivity leads to an EPT
$\overline{\epsilon}_{ij}$ of a form that permits only a single refractive index with a
positive real part. In the following section, we will calculate the
propagation modes to get a definite prediction.

A parity transformation in three spatial dimensions implies
$\cos\theta\mapsto \cos(\pi-\theta)=-\cos\theta$ for the polar angle (and
$\phi\mapsto \pi+\phi$ for the azimuthal angle, which does not occur in the
refractive index).
As the real part of the refractive index in \eqref{eq68} only contains squares of
$\cos\theta$, it is invariant under parity transformations. However, at least the
imaginary part of $n$ exhibits parity-violating properties.

We note in passing that there is a direct connection between the configuration under
consideration and the material studied in Ref.~\cite{Kaushik2}. We call attention to Eq.~(2) in the
latter paper that describes a magnetic conductivity for the material TaAs. This means that the
authors of Ref.~\cite{Kaushik2} have found a microscopic realization of a crystal that leads to
a macroscopic, effective magnetic conductivity of the form of our Eq.~(\ref{eq65}) (with the
identification $\sigma_B\hat{c}_i=-b_i$ with their $\sigma_B$ and~$\hat{c}$).

Our understanding of their result is as follows. They employed a sample-based
coordinate system $(a,b,c)$ and the crystal lattice of TaAs has symmetries with respect to certain
axes and planes of this coordinate system. There is a second-rank tensor $\sigma_P$ that links
components of the current $\mathbf{J}$ and angular momentum $\mathbf{L}$: $J^i=(\sigma_P)^i_{\phantom{i}k}L^k$.\footnote{Note that we employ symbols different from those in Ref.~\cite{Kaushik2} to avoid confusion with some of our quantities.}
The nonzero components of $\sigma_P$ are determined from the symmetry properties of the crystal. The current
$\mathbf{J}$ transforms like a vector under parity transformations and reflections at single planes,
whereas the angular momentum $\mathbf{L}$ transforms like a pseudovector.

Let us consider the $b$-$c$ plane. The above relation links components of the current
and angular momentum in this plane: $J^b=(\sigma_P)^b_{\phantom{b}c}L^c$. If there is a reflection symmetry with
respect to the $b$-$c$ plane, the vectors transform according to the set of (un-numbered) transformation
rules given under Eq.~(1) in Ref.~\cite{Kaushik2}. If $J^b$ is then linked to $L^c$ by a nonzero $(\sigma_P)^b_{\phantom{b}c}$,
there will be a contradiction, as the components of $\mathbf{J}$ in the plane transform differently from the
components of $\mathbf{L}$ in the same plane. So the corresponding component of the tensor $\sigma_P$
must be zero. These arguments imply a magnetic conductivity of TaAs in the form of Eq.~(\ref{eq65}).

\subsubsection{Dielectric nonconducting medium}

Let us come back to Eq.~(\ref{eq68}).
If we start from a dielectric with zero Ohmic conductivity
$(\epsilon \neq 0,\sigma=0,\sigma^{B}\neq 0)$, the latter equations reduce to
\begin{equation}
n=\sqrt{2\Upsilon_b}+{\mathrm i}\Xi_b\,, \quad \Xi_b=\frac{\mu b_3}{2 \omega}\,,
\label{eq69B}
\end{equation}
which is compatible with the behavior of a conducting medium. Therefore, the
off-diagonal chiral conductivity of \eqref{eq65} ascribes a conducting behavior
to the material even for a purely dielectric substrate $(\epsilon \neq 0,\sigma=0)$.

\subsubsection{\label{antisymmetric-modes}Propagation modes}

For a phenomenological analysis of the physics, let us choose a convenient
coordinate system, without loss of generality. Since $\mathbf{n}$
and $\mathbf{b}$ define a plane, to be labeled the $y$-$z$ plane, we take the
$z$ axis along the direction of $\mathbf{n}$ so that
\begin{equation}
\mathbf{n}=(0,0,n)\,,\quad \mathbf{b}=b(0,\sin\theta,\cos\theta)\equiv (0,b_2,b_3)\,.
\label{COORD}
\end{equation}
Recall that $\theta$ is the angle between $\mathbf{n}$ and $\mathbf{b}$.
This choice of coordinates leads to a very simple expression for the matrix
of~\eqref{eq66-1}:
\begin{equation}
\lbrack M_{ij}]=\left(
\begin{array}{ccc}
n^{2}-\mu\epsilon - n \frac{ \mathrm{i}\mu  }{\omega}b_3 & 0 & 0 \\
0 & n^{2}-\mu\epsilon - n \frac{\mathrm{i}\mu }{\omega}b_3 & 0 \\
0 & n \frac{\mathrm{i} \mu  }{\omega}b_2 & -\mu\epsilon \\
\end{array}
\right)\,,
\label{MATRIX}
\end{equation}
which immediately yields the dispersion equation
\begin{equation}
\Big(n^{2}-\mu\epsilon-n  \frac{ \mathrm{i} \mu  }{\omega}b_3\Big)^{2}=0\,,
\label{DISP_EQ}
\end{equation}
with $n b_3=\mathbf{n\cdot b}$. The latter corresponds exactly to \eqref{eq67B}
when $\sigma=0$. Equation~(\ref{DISP_EQ}) has a single solution with a non-negative
real part, given by \eqref{eq69B}.

Recalling that $\sqrt{2\Upsilon_b}=\sqrt{\mu\epsilon-\Xi_b^2}$, we distinguish
between two cases according to the choice of the sign inside the square root of the
above relation. In the first case, when $\mu \epsilon\leq \Xi_{b}^2$, the refractive
index is purely imaginary with a positive imaginary part, which damps propagation.
The alternative, $\mu\epsilon>\Xi_{b}^2$, yields propagation modes
that are to be described now. The condition $M_{ij}E^j=0$ requires
\begin{equation}
E^3=n\frac{\mathrm{i}b_2}{\epsilon\omega}E^2\,,
\label{PROP_COND}
\end{equation}
leaving $E^1$ completely arbitrary. We take advantage of this freedom to choose two
orthogonal vectors satisfying the condition in \eqref{PROP_COND}. We find
\begin{subequations}
\label{modesantisymmetric0}
\begin{align}
\mathbf{E}_{\pm}&=\frac{1}{\sqrt{2(1+Q^2)}}\begin{pmatrix}
\pm\sqrt{1+Q^2} \\
-1 \\
-\mathrm{i} Q e^{\mathrm{i}\alpha}\\
\end{pmatrix}\,, \\[2ex]
Q&=\frac{b_2N}{\epsilon \omega}\,.
\label{modesantisymmetric}
\end{align}
\end{subequations}
Here we parametrized the complex refractive index as
\begin{subequations}
\begin{align}
n&=Ne^{\mathrm{i}\alpha}\,,\quad N=\sqrt{n^{*}n}=\sqrt{\mu\epsilon}\,, \\[2ex]
\tan \alpha&=\frac{\Xi_b}{\sqrt{\mu \epsilon-(\Xi_b)^2}}\,.
\end{align}
\end{subequations}
We can easily verify that $\mathbf{E}_+^* \cdot\mathbf{E}_-=0$ and also that the $y$ and
$z$ components of the propagation modes obey \eqref{PROP_COND}. Thus, we have
recovered two orthogonal modes whose propagation is described by the
same refractive index. This is an unexpected result, which nevertheless is analogous to
the simplest isotropic case (without the presence of a magnetic conductivity)
where two linear polarization modes associated with the same refractive
index occur. Notice that we have $\mathbf{k} \cdot \mathbf{E}_{\pm} \neq 0$
in this case, which prevents  an interpretation of the fields in terms of standard elliptical
polarizations with components defined in the two-dimensional subspace orthogonal to $\mathbf{k}$.

Since in the coordinate system defined in \eqref{COORD} we only have access to the inversion
$z \mapsto -z$, the parity transformation $\mathbf{n} \mapsto -\mathbf{n} $ can be better
studied in a rotated frame where now the vector $\mathbf{b}$ defines the new $z$ axis endowing the
system with axial symmetry. The new frame is obtained via a rotation of the former one by an angle
$\theta=\arccos(b_3/|\mathbf{b}|)$ with respect to an axis perpendicular to the $\mathbf{n}$-$\mathbf{b}$
plane. The latter can be associated with an arbitrary plane having a constant azimuthal angle $\phi$
in spherical coordinates. Calling ${\bar E}^i$ the components of the electric field in the rotated
frame, we derive the following expressions:
\begin{subequations}
\label{modesantisymmetricrotated0}
\begin{align}
\bar{E}_{\pm}^1&=\pm \frac{1}{\sqrt{2}}\,, \displaybreak[0]\\[2ex]
\bar{E}_{\pm}^2&=-\frac{1}{\sqrt{2(1+\bar{Q}^2\sin^2\theta)}}\left[\cos\theta-\mathrm{i}\bar{Q}e^{\mathrm{i}\alpha}\sin^2\theta\right]\,, \displaybreak[0]\\[2ex]
\bar{E}_{\pm}^3&=-\frac{1}{\sqrt{2(1+\bar{Q}^2\sin^2\theta)}}\sin\theta\left[1+\mathrm{i}\bar{Q}e^{\mathrm{i}\alpha}\cos\theta\right]\,,
\end{align}
with
\begin{equation}
\bar{Q}=\frac{bN}{\epsilon\omega}\,.
\end{equation}
\end{subequations}
The magnetic fields corresponding to \eqref{modesantisymmetric0} are obtained using $\mathbf{B}_{\pm}=\mathbf{n}\times
\mathbf{E}_{\pm}$, and those associated with \eqref{modesantisymmetricrotated0} are computed in an analog way.

As $\mathbf{n}$ lies in a plane, we restrict the parity transformation to the replacement rule of the
polar angle: $\theta \mapsto \pi-\theta$. Then, a parity transformation of a vector in two dimensions
corresponds to its reflection at one of the two axes modulo a rotation by $\pi$. Thus, the expected behavior
under a parity transformation in the plane is that only one of the two vector components flips its sign.
The vector $\mathbf{n}$ is transformed in a way that the $n_3$ component is reflected at the $n_2$ axis.
If parity was conserved, the sign of the second components $\bar{E}_{\pm}^2$ should change under a parity
transformation, while the third components $\bar{E}_{\pm}^3$ should remain invariant. However, one can
verify that $\bar{E}_{\pm}^2$ and $\bar{E}_{\pm}^3$ do not behave in this manner. On the contrary, the
terms involving $b$ spoil the behavior expected, which is a signal of parity violation. Note that the
refractive indices of Eqs.~(\ref{eq67a}) and (\ref{eq69B}) also change under parity transformations, as
$\Xi_b\mapsto -\Xi_b$.\\

\subsection{Nondiagonal symmetric conductivity tensor}
\label{sec:nondiagonal-symmetric-configuration}

Now we examine the case where the magnetic conductivity is given by a traceless
symmetric tensor, in accordance with the following parametrization:
\begin{equation}
\sigma_{ij}^{B}=\frac{1}{2}(a_{i}c_{j}+a_{j}c_{i})\,,
\label{eq68-1}
\end{equation}%
where $a_{i}$ and $c_{i}$ are the components of two orthogonal background
vectors $\mathbf{a}$ and $\mathbf{c}$, i.e., $\mathbf{a} \cdot \mathbf{c}=0$
such that $\sigma_{ii}^{B}=0$. Inserting \eqref{eq68-1} into \eqref{eq22} yields
\begin{equation}
\bar{\epsilon}_{ij}=\left( \epsilon +\mathrm{i}{\frac{\sigma }{\omega }}\right) {%
\delta}_{ij}+{\frac{\mathrm{i}}{{2{\omega }^{2}}}}\left(
a_{i}c_{n}+a_{n}c_{i}\right) {\epsilon }_{nbj}k_{b}\,. \label{eq68-2}
\end{equation}\\
The tensor stated in \eqref{eq36B} is explicitly represented by the following matrix:
\begin{subequations}
\begin{widetext}
\begin{align}
\label{eq68-3}
[M_{ij}]&=\begin{pmatrix}
n_2^2+n_3^2-\mu\epsilon & -n_1n_2 & -n_1n_3 \\
-n_1n_2 & n_1^2+n_3^2-\mu\epsilon & -n_2n_3 \\
-n_1n_3 & -n_2n_3 & n_1^2+n_2^2-\mu\epsilon \\
\end{pmatrix} \notag \\
&\phantom{{}={}}-\mathrm{i}\frac{\mu}{2\omega}\begin{pmatrix}
2\sigma+\epsilon_{11} & n_1(a_1c_3+a_3c_1)-2n_3a_1c_1 & -n_1(a_1c_2+a_2c_1)+2n_2a_1c_1 \\
-n_2(a_2c_3+a_3c_2)+2n_3a_2c_2 & 2\sigma+\epsilon_{22} & n_2(a_2c_1+a_1c_2)-2n_1a_2c_2 \\
n_3(a_3c_2+a_2c_3)-2n_2a_3c_3 & -n_3(a_3c_1+a_1c_3)+2n_1a_3c_3 & 2\sigma+\epsilon_{33} \\
\end{pmatrix}\,,
\end{align}
\end{widetext}
where
\begin{align}
\epsilon_{11}& =(a_{1}c_{2}+a_{2}c_{1})n_{3}-(a_{1}c_{3}+a_{3}c_{1})n_{2}\,,
\label{eq68-4} \displaybreak[0]\\[2ex]
\epsilon_{22}& =(a_{2}c_{3}+a_{3}c_{2})n_{1}-(a_{1}c_{2}+a_{2}c_{1})n_{3}\,,
\label{eq68-5} \displaybreak[0]\\[2ex]
\epsilon_{33}& =(a_{3}c_{1}+a_{1}c_{3})n_{2}-(a_{3}c_{2}+a_{2}c_{3})n_{1}\,.
\label{eq68-6}
\end{align}%
\end{subequations}
The evaluation of $\det[M_{ij}]=0$ yields the dispersion equation stated as follows:
\begin{align}
0&=\left[n^2-\mu\left(\epsilon +\mathrm{i}\frac{\sigma}{\omega}\right)+\mathrm{i}\frac{\mu
}{2\omega}\mathbf{n}\cdot(\mathbf{a}\times \mathbf{c})\right] \notag \\
&\phantom{{}={}}\times\left[n^2-\mu\left(\epsilon +\mathrm{i}\frac{\sigma}{\omega}\right)-\mathrm{i}\frac{\mu
}{2\omega}\mathbf{n}\cdot(\mathbf{a}\times \mathbf{c})\right]\,.
\label{eq68-8}
\end{align}
In contrast to the antisymmetric magnetic conductivity of \eqref{eq65}, the new
configuration of \eqref{eq68-1} does not imply a single doubly degenerate refractive index
[cf.~\eqref{eq67B}]. In contrast, we obtain two distinct refractive indices in the current scenario.
Using $\mathbf{n}\cdot (\mathbf{{a}\times {c}})=n|\mathbf{a}||\mathbf{c}|\cos\varphi$ in
\eqref{eq68-8} results in
\begin{subequations}
\label{eq68-10}
\begin{align}
n_{\pm}&=\alpha^{\prime\prime}_+ + \mathrm{i}(\alpha^{\prime\prime}_-\pm \Xi_{a,c})\,, \displaybreak[0]\\[2ex]
\Xi_{a,c}&=\frac{\mu}{{4\omega}}|\mathbf{a}||\mathbf{c}|\cos\varphi\,, \displaybreak[0]\\[2ex]
\alpha^{\prime\prime}_{\pm}&=\sqrt{\sqrt{\Upsilon_{a,c}^2+\left(\frac{\mu\sigma}{2\omega}\right)^{2}}\pm \Upsilon_{a,c}}\,, \displaybreak[0]\\[2ex]
\label{eq:quantity-upsilon-nondiagonal-symmetric}
2\Upsilon_{a,c}&=\mu\epsilon-\Xi_{a,c}^2\,,
\end{align}
\end{subequations}
with the presence of new imaginary terms stemming from the exotic
conductivity and modifying the absorption coefficient.
What is analogous to the antisymmetric configuration
of \eqref{eq65} is the structure of a single one of the two refractive
indices present, i.e., the dependence of the refractive index on the
angle between $\mathbf{k}$ and a three-vector ($\mathbf{b}$
for the antisymmetric case and $\mathbf{a}\times\mathbf{c}$ for the
current scenario). Also, there is again a relative minus sign between
the two contributions in \eqref{eq:quantity-upsilon-nondiagonal-symmetric}.
Therefore, we assume $\mu\epsilon\geq\Xi_{a,c}^2$.

A crucial difference is that two angles play a role
for the current configuration: the angle $\varphi$ between $\mathbf{k}$
and $\mathbf{a}\times\mathbf{c}$ for one mode and the complementary
angle $\pi-\varphi$ for the other mode. This also means that both
modes interchange their roles when $\varphi$ exceeds $\pi/2$. As the
modes differ in their imaginary parts only, birefringence does not
occur. It is merely the attenuation that differs for both modes.

\subsubsection{Dielectric nonconducting medium}

In this case we start from a medium with zero Ohmic conductivity
$(\epsilon \neq 0,\sigma =0,\sigma^{B}\neq 0)$.
Equation~(\ref{eq68-10}) is then reduced to
\begin{equation}
n_{\pm}=\sqrt{2\Upsilon_{a,c}}\pm\mathrm{i}\frac{\mu}{{4\omega}}|\mathbf{a}||\mathbf{c}|\cos\varphi\,,
\label{eq68-11B}
\end{equation}
which exhibits an exotic absorbing behavior for nonconductive
matter, with attenuation coefficient $2\tilde{\alpha}={\mu}|\mathbf{{a}\times {c}}|\cos \varphi$ provided $\mu\epsilon\geq\Xi_{a,c}^2$.
In this case, the modification is proportional to $|\mathbf{{a}\times {c}}|=|\mathbf{a}||\mathbf{c}|$,
which is why such an effect is associated with the nondiagonal elements of
${\sigma }_{ij}^{B}$ instead of its trace.

Note that this nondiagonal magnetic conductivity also provides a conducting behavior
for a dielectric medium, in much the same way as observed for the nondiagonal antisymmetric
case [cf.~\eqref{eq69B}].

\subsubsection{\label{propagationSymmetric} Propagation modes}

In order to examine the propagating modes, we will rewrite the matrix given in \eqref{eq68-3}
for a simplified coordinate system where $\mathbf{a}=(0,a,0)$ and $\mathbf{c}=(0,0,c)$, that is
\begin{subequations}
\begin{align}
M_{ij}&=A\delta_{ij}-n_{i}n_{j} \notag \\
&\phantom{{}={}}+C(\epsilon_{klj}\delta_{i2}\delta_{k3}n_{l}+\epsilon_{klj}\delta_{i3}\delta_{k2}n_{l})\,,
\label{matrix3}
\end{align}
with
\begin{equation}
A=n^{2}-\mu\epsilon \,,\quad C=-\mathrm{i}{%
\frac{\mu}{{2\omega}}}ac\,.
\label{matrix4}
\end{equation}
\end{subequations}
The explicit form of the matrix in \eqref{eq68-3} simplifies to
\begin{align}
\lbrack M_{ij}]&=\begin{pmatrix}
A-n_{1}^{2} & -n_{1}n_{2} & -n_{1}n_{3} \\
-n_{1}n_{2} & A-n_{2}^{2} & -n_{2}n_{3} \\
-n_{1}n_{3} & -n_{2}n_{3} & A-n_{3}^{2} \\
\end{pmatrix} \notag \\
&\phantom{{}={}}+C\begin{pmatrix}
0 & 0 & 0 \\
-n_2 & n_1 & 0 \\
n_3 & 0 & -n_1 \\
\end{pmatrix}\,,
\label{matrix7}
\end{align}
providing the following dispersion equation:
\begin{equation}
(A^{2}-C^{2}n_{1}^{2})(A-n^{2})=0.  \label{matrix8}
\end{equation}
Since $A-n^{2}=-\mu \epsilon$, the dispersion
relations are
\begin{equation}
A=\pm Cn_{1}\,,
\label{matrix9}
\end{equation}
with $n_1=n\cos\varphi$ and corresponding to the $\sigma\mapsto 0$ limit of Eq.~(\ref{eq68-8}).
We observe that the refractive indices depend on the direction of $\mathbf{n}$, which is defined in
terms of the spherical angles $\theta$ and $\phi$ according to \eqref{d15}. Recalling that $\varphi$
is the angle between $\mathbf{n}$ and $\mathbf{a \times c}$ we have that $\cos\varphi=\sin\theta\cos\phi$.

Taking the plus sign in \eqref{matrix9} and using \eqref{matrix7} we obtain
\begin{equation}
E_{+}^{y}={\frac{n_{2}}{n_{1}}}E_{+}^{x}\,,\quad E_{+}^{z}=\left(\frac{{%
Cn_{1}-n_{1}^{2}-n_{2}^{2}}}{{n_{1}n_{3}}}\right) E_{+}^{x}\,.
\label{matrix15}
\end{equation}
Therefore, the electric field for the plus propagating mode is
\begin{equation}
\mathbf{E}_{+}=E_{+}^{(0)}(n_1n_{3},n_{2}n_{3},C n_{1}-n_{1}^{2}-n_{2}^{2})\,,
\label{modesymmetric1}
\end{equation}
with an appropriately chosen amplitude $E_{+}^{(0)}$.
For the negative sign in the dispersion relation of \eqref{matrix9}, we obtain
\begin{equation}
\label{modesymmetric2-0}
E_{-}^{x}=\frac{n_{1}}{n_{3}}E_{-}^{z}\,, \quad E_{-}^{y}=-\frac{Cn_{1}+n_{1}^{2}+n_{3}^{2}}{n_{2}n_{3}}E_{-}^{z}\,.
\end{equation}
Thus, the electric field reads
\begin{equation}
\mathbf{E}_{-}=E_{-}^{(0)}(n_{1}n_2,-(C n_{1}+n_{1}^{2}+n_{3}^{2}),n_2n_{3})\,,
\label{modesymmetric2}
\end{equation}
with another amplitude $E_{-}^{(0)}$.
We observe that $\mathbf{n} \cdot \mathbf{E}_{\pm} \neq 0$.
Equations~(\ref{modesymmetric1}) and (\ref{modesymmetric2}) represent
the propagation modes for the case of a
symmetric exotic magnetic conductivity with the corresponding magnetic fields
given by $\mathbf{B}_{\pm}=\mathbf{n}\times\mathbf{E}_{\pm}$.

\section{\label{section4}Consistency of Maxwell's equations}

For completeness, in  this section we extend the current $\mathbf{J}$ in \eqref{MINH} to include an
external-source contribution $\mathbf{J}_{\mathrm{e}}$ such that now
$\mathbf{J}= \mathbf{J}_{\mathrm{e}}+ \sigma\cdot\mathbf{E}+\sigma^B\cdot\mathbf{B}$. In the previous sections, we
discussed wave propagation outside of sources, which effectively meant to taking $\mathbf{J}_{\mathrm{e}}=0$.
Moreover, the different scenarios we have considered were defined by fixing the current $\mathbf{J}$ via
specific choices of the  electric and magnetic conductivities. Also, the propagation properties of the
fields were obtained just by using Faraday's and Amp\`{e}re's laws, incorporated into \eqref{eq21}, with no
reference to Gauss' law given by the first of Eq.~(\ref{MINH}).

A natural question that arises is the identification of the particular contributions to the charge density
$\rho$ that are consistent with the arbitrary choice of currents, in such a way that charge conservation
${\partial}_{t}{\rho}+\nabla\cdot\mathbf{J}=0$ is preserved. To this end, we work in momentum space with
the standard conventions $\nabla \mapsto \mathrm{i} \mathbf{k}$ and $\partial_t \mapsto -\mathrm{i} \omega$.
The main point to recognize is that Amp\`{e}re's law yields
\begin{equation}
\mathrm{i} \omega \mathbf{k} \cdot \mathbf{D}- \mathbf{k} \cdot \mathbf{J}=0\,,
\end{equation}
which together with Gauss' law $\mathrm{i} \mathbf{k} \cdot \mathbf{D}=\rho$ gives the identification
\begin{equation}
\rho=\mathbf{k} \cdot \mathbf{J}/\omega\,,\label{CHD}
\end{equation}
for $\omega \neq 0$, which is precisely the charge conservation condition in momentum space.
In what follows, we summarize the expressions for the charge densities corresponding to our previous choices
of the magnetic conductivity, setting $\sigma_{ij}=0$, and  recalling that the Maxwell equations retain the
general form of Eqs.~(\ref{MINH}) and (\ref{MHOM}).

For the isotropic case of Sec.~\ref{particular2} with $\mathbf{J}=\mathbf{J}_{\mathrm{e}}+ \Sigma \,\mathbf{B}$,
the charge density is just $\rho=\rho_{\mathrm{e}}=\mathbf{k} \cdot \mathbf{J}_{\mathrm{e}}/\omega$, since
$\mathbf{k} \cdot \mathbf{B}=0$.
The antisymmetric case studied in Sec.~\ref{sec:nondiagonal-antisymmetric-configuration} with
$\mathbf{J}= \mathbf{J}_{\mathrm{e}}- \mathbf{b}\times \mathbf{B}$ yields
\begin{equation}
\rho=\rho_{\mathrm{e}}-\mu\epsilon\mathbf{E}\cdot\mathbf{b}-{\frac{{%
\mathrm{i}\mu}}{\omega}}\mathbf{J}_{\mathrm{e}}\cdot\mathbf{b}\,,
\end{equation}
where we used some of the Maxwell equations in order to get rid of the spatial derivatives arising from \eqref{CHD}.

Finally, the symmetric case of Sec.~\ref{sec:nondiagonal-symmetric-configuration}, where we have
assumed ${\bf{a}}\cdot {\bf{c}}=0$, starts from
\begin{align}
\mathbf{J}&=\mathbf{J}_{\mathrm{e}}+{\frac{1}{2}}[\mathbf{a}(\mathbf{c}\cdot \mathbf{B})+\mathbf{c}(\mathbf{a}\cdot \mathbf{B})] \notag \\
&=\mathbf{J}_{\mathrm{e}}+\mathbf{a}\times (\mathbf{c}\times
\mathbf{B})-{\frac{1}{2}}(\mathbf{a}\times \mathbf{c})\times \mathbf{B}\,,
\label{symm41}
\end{align}
and implies
\begin{align}
\rho&=\rho_{\mathrm{e}}-{\frac{\mu\epsilon}{2}}\mathbf{E}\cdot (\mathbf{a}%
\times \mathbf{c})+{\frac{1}{\omega }}\mathbf{k}\cdot \lbrack \mathbf{a}\times (\mathbf{c}%
\times \mathbf{B})] \notag \\
&\phantom{{}={}}-\frac{\mathrm{i}\mu}{2\omega}{\bf{J}}_{e}\cdot \left({\bf{a}}\times {\bf{c}}\right)\,.
\end{align}
The alternative second form of Eq.~(\ref{symm41}) was motivated by an attempt to
use the remaining Maxwell equations in favor of rewriting \eqref{CHD} without spatial derivatives,
which, unfortunately, was not possible in this case.

\section{\label{section5}Some classical effects}

Towards the end of the paper, we intend to understand the impact that a magnetic conductivity has on
certain phenomena in electrodynamics that are a consequence of material parameters such as refractive
indices. As modified refractive indices for particular choices of a magnetic conductivity have already
been determined earlier, we will now benefit from these findings.

\subsection{\label{section4-1}Skin depth effect}

When an electromagnetic wave falls on the surface of a conductor,
its amplitude will partially penetrate the material due to the attenuation coefficient, while
another part will be reflected. The characteristic penetration length into the
conducting medium defines the skin depth~\cite{Zangwill,Jackson}:
\begin{equation}  \label{skin1}
\bar\delta=\frac{1}{{\omega\,\mathrm{Im}[n]}}\,,
\end{equation}
where $\mathrm{Im}[n]$ is the imaginary part of the complex
refractive index. In the usual scenario for a simple conductor,
the general skin depth reads
\begin{equation}  \label{skin2}
\bar \delta(\omega)={\frac{1}{{\omega n^{\prime\prime}}}}=\sqrt {\frac{2}{{%
\mu\omega\sigma}}}\,,
\end{equation}
for a good (Ohmic) conductor and with $n^{\prime\prime}$ given by %
\eqref{eq14}. Consequently, the skin depth decreases for high frequencies.

We can now write down the skin depth for the particular symmetric (S) and
antisymmetric (AS) scenarios of Sec.~\ref{sec:nondiagonal-antisymmetric-configuration}
and Sec.~\ref{sec:nondiagonal-symmetric-configuration}, respectively, in the case
when the conducting behavior is directly
associated with the magnetic conductivity only, i.e., when $\sigma=0$.
From Eqs.~(\ref{eq69B}) and (\ref{eq68-11B}), we arrive at
\begin{subequations}
\begin{align}
\label{skin6}
\bar \delta_{\mathit{AS}}&={\frac{2}{{\mu b\cos\theta}}}\,, \\
\label{skin7}
\bar \delta_{S}&={\frac{4}{{\mu|\mathbf{{a}\times{c}}|\cos\varphi}}}\,.
\end{align}
\end{subequations}
Therefore, such a skin depth effect does not exhibit a frequency dependence,
which means that the penetration length is the same for all frequency bands.
This is an unusual characteristic for conductors.

\subsection{\label{section4-2}Reflection coefficient at the surface of
conducting matter}

Consider a system composed of an ordinary dielectric characterized by
a refractive index $n_{1}=\sqrt{{\mu}_{1}{\epsilon}_{1}}$ and a
conducting phase of matter with a complex refractive index $n_{2}=n_{2}^{\prime}+\mathrm{i}
n_{2}^{\prime\prime}$ described by the parameters $\epsilon_{2}$, $%
\mu_{2}$ and the Ohmic conductivity $\sigma$. For a wave that propagates
from the dielectric and enters the surface of the conductor, the
reflection coefficient for normal incidence is given by \cite{Zangwill}
\begin{equation}
\label{reflect1}
R=\left| {\frac{{{\mu}_{1}n_{2}^{\prime}-{\mu}_{2}n_{1}+\mathrm{i}{\mu}%
_{1}n_{2}^{\prime\prime}}}{{{\mu}_{1}n_{2}^{\prime}+{\mu}_{2}n_{1}+\mathrm{i}{\mu}%
_{1}n_{2}^{\prime\prime}}}}\right|^{2}\,.
\end{equation}
Considering $n_{1}\ll n_{2}^{\prime}$ one can rewrite \eqref{reflect1} in the
general form
\begin{equation}  \label{reflect2}
R\approx 1-4\left(\frac{{\mu}_{2}}{{\mu}_{1}}\right) {\frac{{n_{1}n_{2}^{\prime}}%
}{{n_{2}^{\prime2}+n_{2}^{\prime\prime2}}}}\,.
\end{equation}
In standard electrodynamics, for a good conductor ($\sigma/(\omega\mu_2)\gg 1$)
one gets $n_{2}^{\prime}=n_{2}^{\prime\prime}=\sqrt{\mu_{2}\sigma/(2\omega)}$.
Then the reflection coefficient $R$ from \eqref{reflect2} yields
\begin{equation}  \label{reflect3}
R\approx1-2\sqrt{2{\frac{{\mu_{2}\epsilon_{1}\omega}}{{\mu_{1}\sigma}}}}\,.
\end{equation}
Setting $\mu_{1}=\mu_{2}$ we obtain the known Hagen-Rubens formula \cite{Zangwill}:
\begin{equation}  \label{reflect4}
R\approx1-2\sqrt{2{\frac{{\epsilon_{1}\omega}}{\sigma}}}\,.
\end{equation}
Now we will derive the version of the latter relation for dielectric media ($\sigma=0$)
endowed with an exotic magnetic conductivity ${\sigma}_{ij}^{B}$.

In the scenario of an antisymmetric ${\sigma}^{B}_{ij}$, the refractive
index is modified according to \eqref{eq69B}, so that
\begin{equation}  \label{reflect8}
n_{2}^{\prime}=\sqrt{{\mu}_{2}{\epsilon}_{2}-\left(\frac{\mu_2}{2\omega}b\cos\theta\right)^2}\,,\quad n_{2}^{\prime\prime}=\frac{\mu_{2}}{2\omega}b\cos\theta\,.
\end{equation}
By inserting \eqref{reflect8} into \eqref{reflect2}, one obtains
\begin{equation}
\label{reflect10}
R_{\mathit{AS}}\approx 1-4\sqrt{{\frac{{%
\mu_{2}\epsilon_{1}}}{{\mu_{1}\epsilon_{2}}}}} \sqrt{1-\frac{\mu_2}{\epsilon_2}\left(\frac{b\cos\theta}{2\omega}\right)^2}\,,
\end{equation}
for real $n_2'$.
This result is a Hagen-Rubens-like formula for the case when there is
a contribution from the antisymmetric magnetic conductivity only. It is also very different from the
reflection coefficient for an ordinary dielectric,
\begin{equation}
\label{reflect11}
R\approx 1-4\sqrt{{\frac{{\mu_{2}\epsilon_{1}}}{{\mu_{1}\epsilon_{2}}}}}\,,
\end{equation}
since the magnetic conductivity introduces a frequency-dependent term in $R$.
Hence, the exotic conductivity results in a conducting-matter phase
in the limit $\sigma\mapsto 0$.

In the scenario of a symmetric ${\sigma}_{ij}^{B}$, one obtains a
similar result:
\begin{equation}
\label{reflect12}
R_{S}\approx 1-4\sqrt{{\frac{{%
\mu_{2}\epsilon_{1}}}{{\mu_{1}\epsilon_{2}}}}}\sqrt{1-\frac{\mu_2}{\epsilon_2}\left(\frac{|\mathbf{a}||\mathbf{c}|\cos\varphi}{4\omega}\right)^2}\,.
\end{equation}

\section{\label{section6}Final Remarks}

Electrodynamics in matter is well described by the Maxwell equations and the constitutive relations.
Extensions of the Maxwell equations including the possibility of a chiral magnetic current is a topical issue
\cite{Li,Chang,Wurff,Landsteiner,Kaushik,Ruiz,Kharzeev2,Qiu}. In this work, we have extended the
scenario of electric currents generated by magnetic fields by studying some basic classical properties
of a magnetic conductivity implemented into the Maxwell equations through the extension of Ohm's law
given in \eqref{eq20}.

The main purpose was to examine the propagation of electromagnetic waves in dispersive dielectric
materials, paying attention to the conduction properties induced upon non-Ohmic materials ($\sigma$=0).
To this end, we proposed some particular realizations for the magnetic-conductivity tensor
${\sigma}_{ij}^{B}$: (i) a diagonal isotropic and a diagonal anisotropic tensor (which include the
chiral magnetic effect) in Secs.~\ref{particular2} and \ref{diagonal-anisotropic} (ii) a nondiagonal
antisymmetric tensor in Sec.~\ref{sec:nondiagonal-antisymmetric-configuration}, and (iii) a traceless
nondiagonal symmetric tensor in Sec.~\ref{sec:nondiagonal-symmetric-configuration}. Let us point out
that the trace of the magnetic-conductivity tensor ${\sigma}_{ij}^{B}$ is related to the chiral
magnetic effect and that the off-diagonal components of ${\sigma }_{ij}^{B}$ describe generalizations
of such an effect. All these configurations induce parity violation, since the associated current
entering Amp\`{e}re's law is linear in $\mathbf{B}$.

We have verified that a diagonal isotropic tensor ${\sigma^B_{ij}=\Sigma \delta_{ij}}$ modifies the
refractive index of a dispersive dielectric medium yielding two distinct complex values $n_{\pm}$.
Most notably, these results are independent of the propagation direction, implying
what we could call an ``isotropic birefringence.'' In the $\sigma=0$ case, the resulting refractive
indices are real and the propagation modes correspond to left- and right-handed circular polarizations.
As such, these media can be characterized as optically active having a frequency-independent specific
rotatory power $\Delta=-\mu \Sigma/2$. This is a consequence of the EPT being Hermitian and the electric
field being orthogonal to the wave vector $\mathbf{k}$. In this case, parity violation manifests
itself only in the fact that $n_+$ is different from~$n_-$.

The diagonal anisotropic case also exhibits birefringence, but this time the complex refractive indices
are direction dependent. Nevertheless, the functions depend on the squares of the momentum components
$k_i$ for $i=1,2,3$ and are insensitive to parity transformations. Thus, the violation of this symmetry
is again  manifest only in the different values of $n_{\pm}$. The EPT is not-Hermitian and
$\mathbf{k} \cdot \mathbf{E} \neq 0$, in general, thus preventing the description of polarization in terms
of left- and right-handed modes.

Focusing on the $\sigma=0$ case, we find again real refractive indices. In this situation we have
explored in detail the case of propagation along $\mathbf{n}=(n_1,n_2,0)$ showing explicitly that the
eigenvectors of the EPT do not correspond to the polarization modes of the electric field. A further
particular case of propagation along $\mathbf{n}=(0,n_2,0)$ restores the orthogonality between
$\mathbf{k}$ and $\mathbf{E}$ and ends up with elliptically polarized propagation modes.
The case of circular polarization is recovered when taking the isotropic limit $\Sigma_i=\Sigma $, which
also reproduces the previous expressions for the refractive indices. As expected, in this specific
situation the eigenvectors of the EPT and the propagation modes coincide. We remark that both cases of
a diagonal chiral conductivity tensor do not induce a conducting behavior in a non-Ohmic dielectric medium.

We have also examined a nondiagonal (antisymmetric and symmetric) ${\sigma}^B_{ij}$. In these cases,
the off-diagonal components of $\sigma_{ij}^{B}$ provide complex refractive indices even for a vanishing
Ohmic conductivity ($\sigma =0$), which leads to the remarkable behavior of a conducting
phase in the dielectric substrate. The magnetic conductivity in these cases implies nonzero absorption
coefficients that damp the intensity of electromagnetic waves propagating through the medium. In both
cases, the EPT is non-Hermitian, $\mathbf{k}\cdot \mathbf{E} \neq 0$, and the refractive indices are
direction dependent. The latter exhibit contributions that obviously induce parity
violation.

A rather unexpected feature in the context of a parity-violating theory occurs for the antisymmetric case,
which is the absence of birefringence. That is to say, we obtain only one doubly degenerate value for the
refractive index, which nevertheless supports two orthogonal polarization modes, as shown in general when
$\sigma=0$. On the contrary, the symmetric case exhibits birefringence.

For the nondiagonal cases (antisymmetric or symmetric) we have calculated the associated skin depth, which
in the limit $\sigma \mapsto 0$ becomes a constant for all frequency bands of the electromagnetic wave
that enters the medium. We also have derived a generalization of the Hagen-Rubens relation for nonconducting
media ($\sigma=0$), with contributions stemming from the magnetic conductivity only. In this case, the reflection
coefficient is frequency dependent, which is a property that does not occur at the interface of two ordinary
dielectric substrates.

It is interesting to observe that when $\sigma=0$, the equation determining the refractive indices in all
four cases considered boils down to the form
\begin{subequations}
\begin{equation}
n^{2}-\mu\epsilon= n B\,,
\end{equation}%
where $B$ specifies each case according to
\begin{equation}
B=\pm\frac{\mu\Sigma}{\omega}\,,
\end{equation}
for the isotropic diagonal case of Sec.~\ref{particular2},
\begin{equation}
B=\pm\frac{\sqrt{\tilde\Omega}}{\omega}\,,
\end{equation}
for the anisotropic diagonal case of Sec.~\ref{diagonal-anisotropic},
\begin{equation}
B=\mathrm{i}\frac{\mu}{\omega}b\cos\theta\,,
\end{equation}
for the antisymmetric case of Sec.~\ref{sec:nondiagonal-antisymmetric-configuration}
and finally,
\begin{equation}
B=\pm\mathrm{i}\frac{\mu}{2\omega}|\mathbf{a}||\mathbf{c}|\cos\varphi\,,
\end{equation}
\end{subequations}
for the symmetric case of Sec.~\ref{sec:nondiagonal-symmetric-configuration}.

A comment in relation to the consistency of our calculation is now in order. A current density $\mathbf{J}$
linear in the magnetic field has been the only additional input we have introduced into the calculation of
the propagation properties of a wave, without making any statement about the corresponding charge density
required by current conservation. The point to be recalled is that when $\omega \neq 0$, Amp\`{e}re's law
directly yields charge conservation via the use of Gauss' law, as shown in Sec.~\ref{section4}. Thus, for
completeness, in the latter section we have identified the charge densities corresponding to some of the currents
we have introduced before. Perhaps an unexpected feature is that some of the parameters defining such currents
($\mathbf{b}$ and $\mathbf{a}\times \mathbf{c}$, for example) give rise to additional contributions to the charge
density proportional to the external currents that we have set equal to zero in this analysis.

Finally, we conclude that this work presents a classical perspective of some novel effects that a magnetic
conductivity can provide for the propagation of electromagnetic waves in dispersive media.

\section*{Acknowledgments}

The authors thank S.~Kaushik, D.E.~Kharzeev, and E.J.~Philip for valuable discussions as to their
work~\cite{Kaushik2} as well as M.~Kaminski and C.~Valagiannopoulos for additional comments. P.D.S.S., M.M.F., and M.S. express their gratitude to FAPEMA, CNPq and CAPES (Brazilian
research agencies) for invaluable financial support. In particular, M.M.F. is supported by FAPEMA Universal
01187/18, and CNPq Produtividade 311220/2019-3. M.S. receives support from FAPEMA
Universal 01149/17, CNPq Universal 421566/2016-7, and CNPq Produtividade 312201/2018-4. Furthermore, we are
indebted to CAPES/Finance Code 001. L.F.U. acknowledges support from the project DGAPA-UNAM-IN103319.


\begin{thebibliography}{99}

\bibitem{Kharzeev1} D.E.~Kharzeev, The chiral magnetic effect and anomaly-induced transport, \href{https://doi.org/10.1016/j.ppnp.2014.01.002}
{Prog. Part. Nucl. Phys. {\bf 75}, 133 (2014)};
D.E.~Kharzeev, J.~Liao, S.A.~Voloshin, and G.~Wang, Chiral magnetic and vortical effects in high-energy nuclear collisions -- A status report, \href{https://doi.org/10.1016/j.ppnp.2016.01.001}{Prog. Part. Nucl. Phys. \textbf{88}, 1 (2016)};
D.~Kharzeev, K.~Landsteiner, A.~Schmitt and H.U.~Yee, \textit{Strongly Interacting Matter in Magnetic Fields},
Lect. Notes Phys. Vol. \textbf{871} (Springer-Verlag, Berlin, Heidelberg, 2013).

\bibitem{Fukushima} K.~Fukushima, D.E.~Kharzeev, and H.J.~Warringa, Chiral magnetic effect,
\href{https://doi.org/10.1103/PhysRevD.78.074033}{Phys. Rev. D \textbf{78}, 074033 (2008)}.

\bibitem{Gabriele} G.~Inghirami, M.~Mace, Y.~Hirono, L.~Del Zanna, D.E.~Kharzeev, and M.~Bleicher, Magnetic fields in heavy ion collisions: flow and charge transport, \href{https://link.springer.com/article/10.1140/epjc/s10052-020-7847-4}{Eur. Phys. J. C \textbf{80}, 293 (2020)}.

\bibitem{Schober} J.~Schober, A.~Brandenburg and I.~Rogachevskii, Chiral fermion asymmetry in high-energy plasma simulations,
\href{https://doi.org/10.1080/03091929.2019.1591393}{Geophys. Astrophys. Fluid Dyn. \textbf{114}, 106 (2020)}.

\bibitem{Vilenkin} A.~Vilenkin, Equilibrium parity-violating current in a magnetic field,
\href{https://doi.org/10.1103/PhysRevD.22.3080}{Phys. Rev. D \textbf{22}, 3080 (1980)};
A.~Vilenkin and D.A.~Leahy, Parity nonconservation and the origin of cosmic magnetic fields,
\href{https://ui.adsabs.harvard.edu/abs/1982ApJ...254...77V/abstract}{Astrophys. J. \textbf{254}, 77 (1982)}.

\bibitem{Maxim} M.~Dvornikov and V.B.~Semikoz, Influence of the turbulent motion on the chiral magnetic effect in the early universe,
\href{https://doi.org/10.1103/PhysRevD.95.043538}{Phys. Rev. D \textbf{95}, 043538 (2017)}.

\bibitem{Leite} G.~Sigl and N.~Leite, Chiral magnetic effect in protoneutron stars and magnetic field spectral evolution,
\href{https://doi.org/10.1088/1475-7516/2016/01/025}{J. Cosmol. Astropart. Phys. 01 (2016) 025}.

\bibitem{Dvornikov} M.~Dvornikov and V.B.~Semikoz, Magnetic field instability in a neutron star driven by the electroweak electron-nucleon interaction versus the chiral magnetic effect,
\href{https://doi.org/10.1103/PhysRevD.91.061301}{Phys. Rev. D \textbf{91}, 061301(R) (2015)}.

\bibitem{Bubnov} A.F.~Bubnov, N.V.~Gubina, and V.Ch.~Zhukovsky, Vacuum current induced by an axial-vector condensate and electron anomalous magnetic moment in a magnetic field,
\href{https://doi.org/10.1103/PhysRevD.96.016011}{Phys. Rev. D \textbf{96}, 016011 (2017)}.

\bibitem{Akamatsu} Y.~Akamatsu and N.~Yamamoto, Chiral Plasma Instabilities,
\href{https://doi.org/10.1103/PhysRevLett.111.052002}{Phys. Rev. Lett. \textbf{111}, 052002 (2013)};
A.~Boyarsky, O.~Ruchayskiy, and M.~Shaposhnikov, Long-Range Magnetic Fields in the Ground State of the Standard Model Plasma,
\href{https://doi.org/10.1103/PhysRevLett.109.111602}{Phys. Rev. Lett. \textbf{109}, 111602 (2012)}.

\bibitem{Maxim1} M.~Dvornikov and V.B.~Semikoz, Instability of magnetic fields in electroweak plasma driven by neutrino asymmetries,
\href{https://doi.org/10.1088/1475-7516/2014/05/002}{J. Cosmol. Astropart. Phys. 05 (2014) 002};
M.~Dvornikov, Chiral magnetic effect in the presence of an external axial-vector field,
\href{https://doi.org/10.1103/PhysRevD.98.036016}{Phys. Rev. D \textbf{98}, 036016 (2018)}.

\bibitem{Maxim2} M.~Dvornikov, Electric current induced by an external magnetic field in the presence of electroweak matter,
\href{https://doi.org/10.1051/epjconf/201819105008}{EPJ Web Conf. \textbf{191}, 05008 (2018)}.

\bibitem{Burkov} A.A.~Burkov, Chiral anomaly and transport in Weyl metals,
\href{https://doi.org/10.1088/0953-8984/27/11/113201}{J. Phys. Condens. Matter \textbf{27}, 113201 (2015)}.

\bibitem{Li} Q.~Li, D.E.~Kharzeev, C.~Zhang, Y.~Huang, I.~Pletikosi\'{c}, A.V.~Fedorov, R.D.~Zhong, J.A.~Schneeloch, G.D~Gu, and T.~Valla,
Chiral magnetic effect in $\mathrm{ZrTe_5}$,
\href{https://doi.org/10.1038/nphys3648}{Nat. Phys. \textbf{12}, 550 (2016)}.

\bibitem{Chang} M.-C.~Chang and M.-F.~Yang, Chiral magnetic effect in a two-band lattice model of Weyl semimetal,
\href{https://doi.org/10.1103/PhysRevB.91.115203}{Phys. Rev. B \textbf{91}, 115203 (2015)}.

\bibitem{Wurff} E.C.I~van der Wurff and H.T.C.~Stoof, Anisotropic chiral magnetic effect from tilted Weyl cones,
\href{https://doi.org/10.1103/PhysRevB.96.121116}{Phys. Rev. B \textbf{96}, 121116(R) (2017)}.

\bibitem{Landsteiner} K.~Landsteiner, Anomalous transport of Weyl fermions in Weyl semimetals,
\href{https://doi.org/10.1103/PhysRevB.89.075124}{Phys. Rev. B \textbf{89}, 075124 (2014)}.

\bibitem{Kaushik} S.~Kaushik and D.E.~Kharzeev, Quantum oscillations in the chiral magnetic conductivity,
\href{https://doi.org/10.1103/PhysRevB.95.235136}{Phys. Rev. B \textbf{95}, 235136 (2017)}.

\bibitem{Ruiz} A.~Mart\'{i}n-Ruiz, M.~Cambiaso, and L.F.~Urrutia, Electromagnetic fields induced by an electric charge near a Weyl semimetal,
\href{https://doi.org/10.1103/PhysRevB.99.155142}{Phys. Rev. B \textbf{99}, 155142 (2019)}.

\bibitem{Kharzeev2} D.E.~Kharzeev and H.J.~Warringa, Chiral magnetic conductivity,
\href{https://doi.org/10.1103/PhysRevD.80.034028}{Phys. Rev. D \textbf{80}, 034028 (2009)}; D.E.~Kharzeev, Chiral magnetic superconductivity,
\href{https://doi.org/10.1051/epjconf/201713701011}{EPJ Web Conf. \textbf{137}, 01011 (2017)}.

\bibitem{Qiu} Z.~Qiu, G.~Cao and X.-G.~Huang, Electrodynamics of chiral matter,
\href{https://doi.org/10.1103/PhysRevD.95.036002}{Phys. Rev. D \textbf{95}, 036002 (2017)}.

\bibitem{Kostelecky:1988zi} V.A.~Kosteleck\'{y} and S.~Samuel, Spontaneous breaking of Lorentz symmetry in string theory,
\href{https://doi.org/10.1103/PhysRevD.39.683}{Phys. Rev. D \textbf{39}, 683 (1989)};
V.A.~Kosteleck\'{y} and R.~Potting, $CPT$ and strings,
\href{https://doi.org/10.1016/0550-3213(91)90071-5}{Nucl. Phys. B \textbf{359}, 545 (1991)};
V.A.~Kosteleck\'{y} and R.~Potting, $CPT$, strings, and meson factories,
\href{https://doi.org/10.1103/PhysRevD.51.3923}{Phys. Rev. D \textbf{51}, 3923 (1995)}.

\bibitem{Colladay} D.~Colladay and V.A.~Kosteleck\'{y}, $CPT$ violation and the standard model,
\href{https://doi.org/10.1103/PhysRevD.55.6760}{Phys. Rev. D \textbf{55}, 6760 (1997)};
D.~Colladay and V.A.~Kosteleck\'{y}, Lorentz-violating extension of the standard model,
\href{https://doi.org/10.1103/PhysRevD.58.116002}{Phys. Rev. D \textbf{58}, 116002 (1998)};
S.~Coleman and S.L.~Glashow, High-energy tests of Lorentz invariance,
\href{https://doi.org/10.1103/PhysRevD.59.116008}{Phys. Rev. D \textbf{59}, 116008 (1999)}.

\bibitem{CFJ} S.M.~Carroll, G.B.~Field, and R.~Jackiw, Limits on a Lorentz- and parity-violating modification of electrodynamics,
\href{https://doi.org/10.1103/PhysRevD.41.1231}{Phys. Rev. D \textbf{41}, 1231 (1990)};
A.A.~Andrianov and R.~Soldati, Lorentz symmetry breaking in Abelian vector-field models with Wess-Zumino interaction,
\href{https://doi.org/10.1103/PhysRevD.51.5961}{Phys. Rev. D \textbf{51}, 5961 (1995)};
A.A.~Andrianov and R.~Soldati, Patterns of Lorentz symmetry breaking in QED by $CPT$-odd interaction,
\href{https://doi.org/10.1016/S0370-2693(98)00823-5}{Phys. Lett. B \textbf{435}, 449 (1998)};
A.A.~Andrianov, R.~Soldati, and L.~Sorbo, Dynamical Lorentz symmetry breaking from a (3+1)-dimensional axion-Wess-Zumino model,
\href{https://doi.org/10.1103/PhysRevD.59.025002}{Phys. Rev. D \textbf{59}, 025002 (1998)};
J.~Alfaro, A.A.~Andrianov, M.~Cambiaso, P.~Giacconi, and R.~Soldati, Bare and induced Lorentz and $CPT$ invariance violations in QED,
\href{https://doi.org/10.1142/S0217751X10049293}{Int. J. Mod. Phys. A \textbf{25}, 3271 (2010)};
A.A.~Andrianov, D.~Espriu, P.~Giacconi, and R.~Soldati, Anomalous positron excess from Lorentz-violating QED,
\href{https://doi.org/10.1088/1126-6708/2009/09/057}{J. High Energy Phys. 09 (2009) 057}.

\bibitem{KM} V.A.~Kosteleck\'{y} and M.~Mewes, Cosmological Constraints on Lorentz Violation in Electrodynamics,
\href{https://doi.org/10.1103/PhysRevLett.87.251304}{Phys. Rev. Lett. \textbf{87}, 251304 (2001)};
V.A.~Kosteleck\'{y} and M.~Mewes, Signals for Lorentz violation in electrodynamics,
\href{https://doi.org/10.1103/PhysRevD.66.056005}{Phys. Rev. D \textbf{66}, 056005 (2002)};
V.A.~Kosteleck\'{y} and M.~Mewes, Sensitive Polarimetric Search for Relativity Violations in Gamma-Ray Bursts,
\href{https://doi.org/10.1103/PhysRevLett.97.140401}{Phys. Rev. Lett. \textbf{97}, 140401 (2006)};
C.A.~Escobar and M.A.G.~Garcia, Full $CPT$-even photon sector of the standard model extension at finite temperature,
\href{https://doi.org/10.1103/PhysRevD.92.025034}{Phys. Rev. D \textbf{92}, 025034 (2015)};
A.~Mart\'{i}n-Ruiz and C.A.~Escobar, Casimir effect between ponderable media as modeled by the standard model extension,
\href{https://doi.org/10.1103/PhysRevD.94.076010}{Phys. Rev. D \textbf{94}, 076010 (2016)}.

\bibitem{Bailey} Q.G.~Bailey and V.A.~Kosteleck\'{y}, Lorentz-violating electrostatics and magnetostatics,
\href{https://doi.org/10.1103/PhysRevD.70.076006}{Phys. Rev. D \textbf{70}, 076006 (2004)}.

\bibitem{Gurnett} D.A. Gurnett and A. Bhattacharjee, {\it Introduction to Plasma Physics}
(Cambridge University Press, Cambridge, England, 2005).

\bibitem{Palash} J.F.~Nieves and P.B.~Pal, Third electromagnetic constant of an isotropic medium,
\href{https://doi.org/10.1119/1.17598}{Am. J. Phys. \textbf{62}, 207 (1994)}.

\bibitem{Urrutia2} A.~Mart\'{i}n-Ruiz, M.~Cambiaso, and L.F.~Urrutia, Electro- and magnetostatics of topological insulators as modeled by planar, spherical, and cylindrical $\theta$ boundaries: Green's function approach,
\href{https://doi.org/10.1103/PhysRevD.93.045022}{Phys. Rev. D \textbf{93}, 045022 (2016)}.

\bibitem{Zangwill} A.~Zangwill, \textit{Modern Electrodynamics} (Cambridge University Press, New York, 2012).

\bibitem{Kshetrimayum} R.S.~Kshetrimayum, A brief intro to metamaterials, \href{https://ieeexplore.ieee.org/document/1368916}{IEEE Potentials {\bf 23}, 44 (2004)}.

\bibitem{Landau} L.D.~Landau and E.M.~Lifshitz, \textit{Electrodynamics
of Continuous Media, Course of Theoretical Physics, Vol. 8}, 2nd ed. (Pergamon Press, New York, 1984).

\bibitem{Jackson} J.D.~Jackson, \textit{Classical Electrodynamics}, 3rd ed. (John Wiley \& Sons, New York, 1999).

\bibitem{Birefringence1} L.A.~Pajdzik and A.M.~Glazer, Three-dimensional birefringence imaging with a microscope tilting-stage. I. Uniaxial
crystals, \href{https://doi.org/10.1107/S0021889806007758}{J. Appl. Cryst. {\bf 39}, 326 (2006)}.

\bibitem{Birefringence2} I.G.~Wood and A.M.~Glazer, Ferroelastic phase transition in $\mathrm{BiVO_4}$. I. Birefringence measurements using the rotating-analyser method, \href{https://doi.org/10.1107/S002188988001196X}{J. Appl. Cryst. {\bf 13}, 217 (1980)}; M.A. Geday, W. Kaminsky, J.G. Lewis, and A.M. Glazer, Images of absolute retardance $L\cdot\Delta n$, using the rotating polariser method, \href{https://doi.org/10.1046/j.1365-2818.2000.00687.x}{J. Microsc. {\bf 198}, 1 (2000)}.

\bibitem{Kaushik2} S.~Kaushik, D.E.~Kharzeev, and E.J.~Philip, Transverse chiral magnetic photocurrent induced by linearly polarized light in symmetric Weyl semimetals, \href{https://arxiv.org/abs/2006.04857}{arXiv:2006.04857}.

\end{thebibliography}
\end{document}